\documentclass[pre,twocolumn,aps,superscriptaddress,showpacs]{revtex4}

\usepackage{graphicx}
\usepackage{amssymb}
\usepackage{mathrsfs}
\usepackage[british]{babel}
\usepackage{amstext}
\usepackage{color}

\begin{document}

\title{Optimization of random search processes in the presence of an
external bias}

\author{Vladimir V. Palyulin}
\affiliation{Institute for Physics \& Astronomy, University of Potsdam,
D-14476 Potsdam-Golm, Germany}
\author{Aleksei V. Chechkin}
\affiliation{Akhiezer Institute for Theoretical Physics NSC KIPT,
Kharkov 61108, Ukraine}
\affiliation{Max Planck Institute for the Physics of Complex Systems, D-01187
Dresden, Germany}
\author{Ralf Metzler}
\email{rmetzler@uni-potsdam.de}
\affiliation{Institute for Physics \& Astronomy, University of Potsdam,
D-14476 Potsdam-Golm, Germany}
\affiliation{Physics Department, Tampere University of Technology,
FI-33101 Tampere, Finland}

\date{\today}

\begin{abstract}
We study the efficiency of random search processes based on L{\'e}vy flights with
power-law distributed jump lengths in the presence of an external drift, for
instance, an underwater current, an airflow, or simply the bias of the searcher
based on prior experience. While L\'evy flights turn out to be efficient search
processes when relative to the starting point the target is upstream, in the
downstream scenario regular Brownian motion turns out to be advantageous. This is
caused by the occurrence of leapovers of L{\'e}vy flights, due to which L{\'e}vy
flights typically overshoot a point or small interval. Extending our recent
work on biased LF search [V. V. Palyulin, A. V. Chechkin, and R. Metzler, Proc.
Natl. Acad. Sci. USA, \textbf{111}, 2931 (2014).] we establish criteria when
the combination of the external stream and the initial distance between the
starting point and the target favors L{\'e}vy flights over regular Brownian search.
Contrary to the common belief that L{\'e}vy flights with a L{\'e}vy index $\alpha
=1$ (i.e., Cauchy flights) are optimal for sparse targets, we find that the optimal
value for $\alpha$ may range in the entire interval $(1,2)$ and include Brownian
motion as the overall most efficient search strategy.
\end{abstract}


\maketitle

\section{Introduction}  

To find a lost key on a parking lot or a paper on an untidy desk are typical
everyday experiences for search problems \cite{BenichouIntermittent,newbybresloff}.
Search processes occur on many
different scales, ranging from the passive diffusive search of regulatory
proteins for their specific binding site in living biological cells \cite{cells}
over the search of animals for food \cite{nathan,gandhi} or of computer algorithms
for minima in a complex search space \cite{ilya}. Here we are interested in random,
jump-like search processes. The searcher, that is, has no prior information on the
location of its target and performs a random walk until encounter with the target.
During a relocation along its trajectory (a jump), the walker is insensitive to
the target. In the words of movement ecology occupied with the movement patterns
of animals, this process is called \emph{blind search\/} with \emph{saltatory
motion}. It is typical for predators hunting at spatial scales exceeding their
sensory range \cite{james,sharks,humphries,burrows}. For instance, blind search
is observed for plankton-feeding basking sharks \cite{sims2006}, jellyfish
predators and leatherback turtles \cite{hays2006}, and southern elephant seals
\cite{elephantseal}. Saltatory search is distinguished from \emph{cruise search},
when the searcher continues to explore its environment during relocations.

The first studies on random search considered the Brownian motion of the searcher
as a default strategy, until Shlesinger and Klafter proposed that L{\'e}vy flights
(LFs) are much more efficient in the search for sufficiently sparse targets
\cite{shlekla}. In a Markovian LF the individual displacement lengths $x$ of the
walker are power-law distributed, $\lambda(x)\sim|x|^{-1-\alpha}$, where due to
$0<\alpha<2$ the second moment of the jump lengths diverges, $\langle x^2\rangle
\to\infty$ \cite{report}. This lack of a length scale $\langle x^2\rangle^{1/2}$
effects a fractal dimension of the trajectory \cite{hughes,mahsa}, such that
local search is interspersed by long, decorrelating excursions. This strategy
avoids oversampling, the frequent return to previously visited points in space
of recurrent random walk processes, such as Brownian motion in one and two
dimensions \cite{gandhi,shlekla,viswanathan,bartumeus}. The latter are indeed the
relevant cases for land-based searchers. Even for airborne or marine searchers,
the vertical span of their trajectories is usually much smaller than the horizontal
span, rendering their motion almost fully two dimensional. The outstanding role of
LFs for random search processes in one and two dimensions was formulated in the
\emph{LF hypothesis: Superdiffusive motion governed by fat-tailed propagators
optimize encounter rates under specific (but common) circumstances: hence some
species must have evolved mechanisms that exploit these properties $[\ldots]$}
\cite{gandhi}.

Starting with the report of LF-search by albatross birds \cite{viswa1996,
viswanathan} there was a surge of discoveries of such scale-free search
strategies, \emph{inter alia}, for marine predators \cite{Sharks1,Sharks2},
insects such as moths \cite{moths}, land-based mammals such as deer and goats
\cite {deers,goats}, as well as microorganisms such as dinoflagellates
\cite{dinoflagellate}. In some cases these reports were debated. For instance, an
additional investigation showed that spider monkeys indeed move deterministically
\cite{spider} and mussels have multimodal rather than power-law relocations
\cite{mussel1,mussel2}. Similarly, plant lice exhibit L{\'e}vy motion on the
population level but not for the motion of individuals \cite{petrovskii}.
In particular, the disqualification of the LF statistics for albatrosses
\cite{albatros} became a strong argument against the LF hypothesis.
However, there is strong evidence that for individual albatross birds LFs are
indeed a real search pattern \cite{PNAS2012}.

From extensive studies of human trajectories it was shown that LF motion patterns
are indeed characteristic \cite{Brockmann,BrockmannNature}, although in some cases
correlations in the motion exist \cite{tal}. Similarly the paths of fishing boats
follow LF statistics \cite{seiners}, although this observation has also been put
in question \cite{edwards}. Interesting findings were reported from robotics,
where one of the important questions is how robots should search for hidden
targets. From simulations it was concluded that the most successful robots
performed motion consistent with LF foraging \cite{robot2004}.

A number of other search strategies was proposed as an alternative to LFs. These
include intermittent dynamics switching between local diffusive search and
ballistic relocations \cite{intermittent}, which may also be of power-law form
\cite{PNAS2008}. Moreover, searchers may perform persistent random walks with
finite tangential correlations \cite{benhamou}. However, while the difference in
performance between models with a scale and LFs may be small, the central advantage
of the LF strategy is its robustness: while other models work best when their
parameters are optimized for specific environmental conditions such as the target
density LFs remain close to optimal even when these conditions are altered
\cite{PNAS2008}. LFs have thus been promoted as a preferred strategy when there
is insufficient prior knowledge on the search space. In particular, for
sufficiently sparse targets several analyses claim that the optimal value for
the power-law exponent is $\alpha=1$
\cite{viswanathan,PNAS2008,michael,bartumeus1,bartumeus2,reynolds1,reynolds2}.

Following the continuing debate over the validity of the LF hypothesis we here
extend our recent study in Ref.~\cite{pnas} and
scrutinize the LF hypothesis from a different angle. Namely, we analyze the
performance of the LF search mechanism in the presence of an external bias. Such
a bias could, for instance, correspond to an underwater current biasing the
search motion of marine predators or search robots, or to an airflow driving
birds of prey in a preferential direction. It could be a bias in an abstract
landscape searched by a computed algorithm. Finally, it could also simply be
the personal bias of the searcher based on some prior experience. Such biases
are commonplace and their effect needs to be considered in models for random
search processes. It turns out that an external bias may have profound consequences
for the efficiency of LF search. Thus, even in the presence of a small bias LF
search may fare worse than Brownian search. Depending on the initial separation
between searcher and target and on whether from the perspective of the searcher
the bias is directed towards or away from the target, we find that the optimal
power-law exponent $\alpha$ may range in the entire interval from one to 2, and
explicitly include Brownian motion. Without prior knowledge, it may turn out that
Brownian search is indeed the more efficient search method.

The paper is organized as follows. In Sec.~II we set up our model in terms of the
fractional Fokker-Planck equation. In Sec.~III we explicitly calculate the first
arrival density and the search efficiency in absence of a bias. In particular,
we obtain the optimal search parameters as function of the initial distance between
searcher and target. In Sec.~IV we generalize these findings to the case when an
external bias is present. We analyze the temporal decay of the first arrival
density and introduce a generalized P{\'e}clet number. We draw our Conclusions
in Sec.~V. In the Appendix, we detail several calculations.

\section{A Model}

\subsection{Solution of the Fokker-Planck equation
with the sink term}

\begin{figure}
\includegraphics[width=8.6cm]{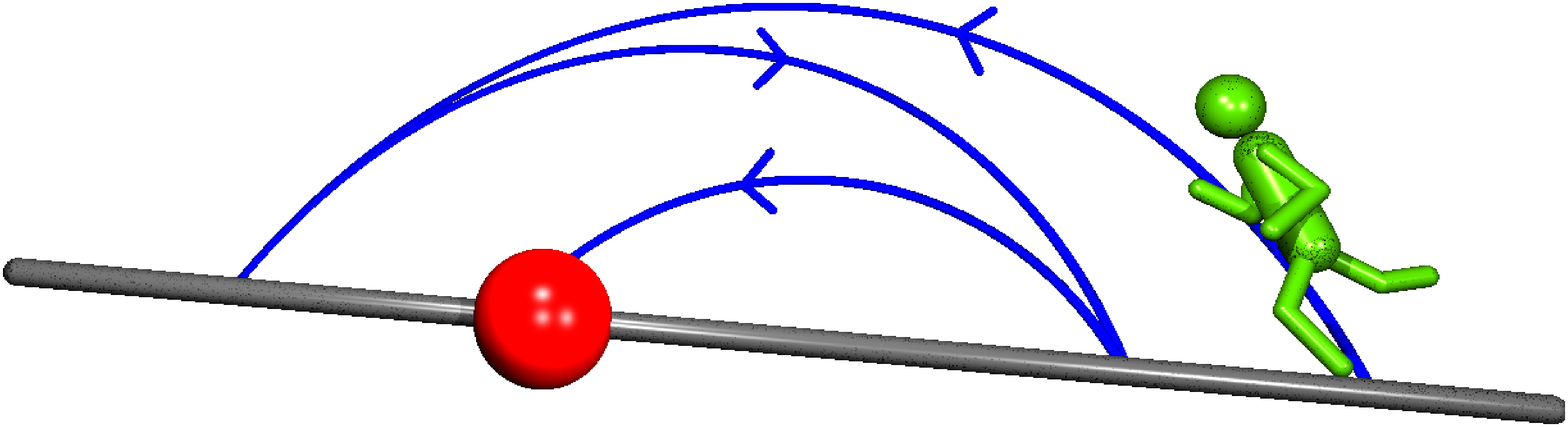}
\includegraphics[width=8.6cm]{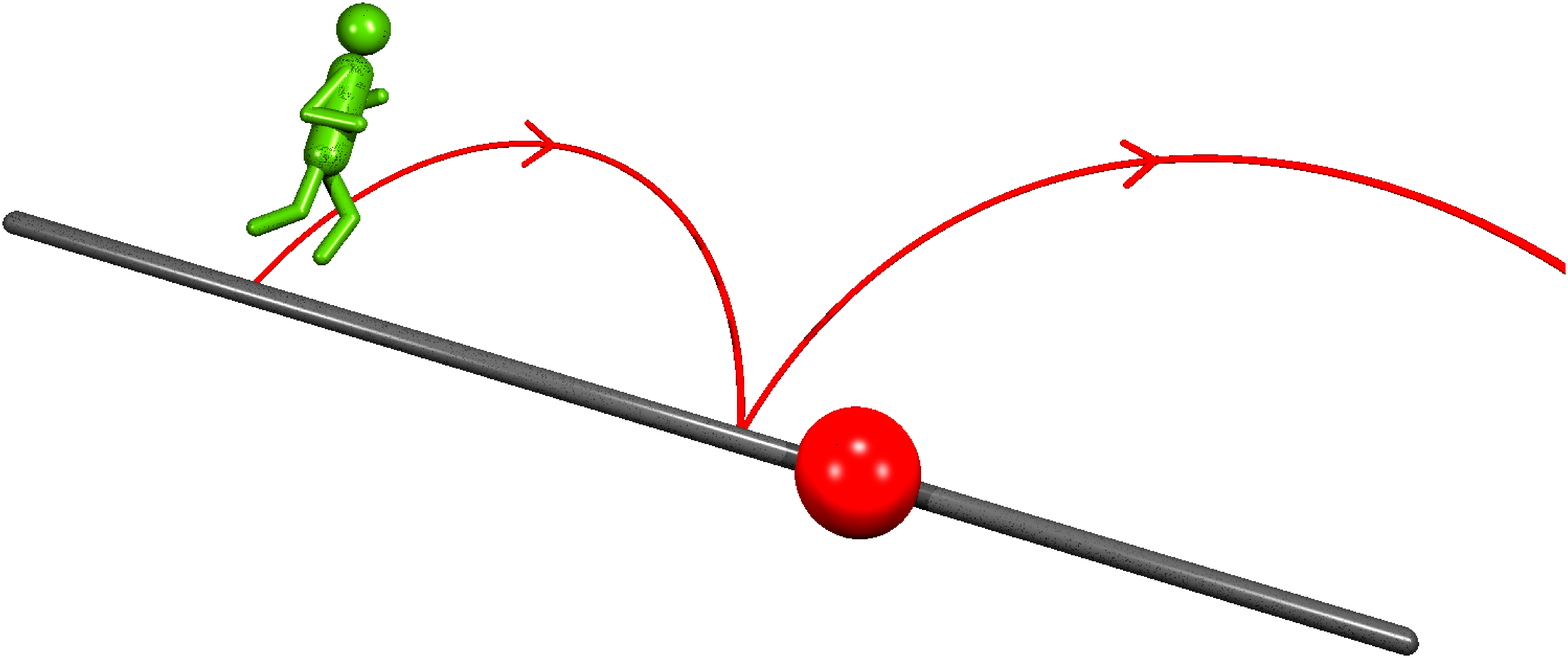}
\caption{Schematic of the search process. A random walker performs random jumps in the search space. Top: the search is initially biased by a drift away from the target. The walker first overshoots twice the target in so-called leapovers before hitting the target. Bottom: the bias is initially directed in the direction of the target. The searcher overshoots the target and is pushed away from the target.
No detection occurs.}
\label{Sketch}
\end{figure}

Imagine the process depicted in Fig.~1. A random walker moves by random jumps,
whose lengths are chosen according to the power-law distribution $\lambda(x)$
with
\begin{equation}
\label{jld}
\lambda(k)=\exp\left(-\sigma^{\alpha}|k|^{\alpha}\right)\,\,\Rightarrow\,\,
\lambda(x)\simeq\frac{\sigma^{\alpha}}{|x|^{1+\alpha}}
\end{equation}
where $\lambda(k)=\int_{-\infty}^{\infty}\lambda(x)\exp(ikx)dx$ denotes the
Fourier transform of $\lambda(x)$. Its stretched Gaussian form in $k$ space
defines a symmetric L{\'e}vy stable distribution, with the long-tailed
asymptotic form $\lambda(x)\simeq|x|^{-1-\alpha}$ for $0<\alpha<2$.
Consequently the variance of $\lambda(x)$ diverges, $\langle x^2\rangle\to\infty$,
while fractional moments $\langle|x|^{\delta}\rangle$ of order $0<\delta<\alpha$
are finite \cite{hughes,report}. In addition to such L{\'e}vy stable jump lengths
we consider an external drift. This \emph{bias\/} is called uphill (or downhill)
when the bias is directed against (along) the walker with respect to the original
walker-target location.

A searcher finds the target when after a jump its position coincides with the
location of the target. That is, the successful search corresponds to the first
arrival at the target coordinate. This process will be substantially different
from that of Brownian search, and consists of a tradeoff between two effects:
(i) due to the scale-free jump length distribution (\ref{jld}) the above-mentioned
oversampling is diminished, and less points are revisited multiply before eventual
location of the target. (ii) Due to the existence of extremely long jumps the
walker may severely overshoot the target, producing so-called leapovers (see
Fig.~1). The length of these leapovers is distributed as $\wp_l(\ell)\simeq\ell^{
-1-\alpha/2}$, and is thus wider than the original jump length distribution
(\ref{jld}) \cite{leapover}. This property renders the first arrival of an LF
different from the process of first passage, and the first arrival efficiency
worsens with decreasing $\alpha$ \cite{JPA2003}.

The basis for the mathematical description of the first arrival process of LFs
is the fractional Fokker-Planck equation for the density $f(x,t)$ to find the
searcher at position $x$ at time $t$ in the presence of an external force
field \cite{report,CheGo}. Generalizing the approach of Ref.~\cite{JPA2003}, to
describe the first arrival of a searcher to point $x=0$, we remove the searcher
from the target location $x=0$ by a $\delta$-sink with time dependent weight
$\wp_{\mathrm{fa}}(t)$,
\begin{equation}
\frac{\partial f(x,t)}{\partial t}=K_{\alpha}\frac{\partial^{\alpha}f(x,t)}{
\partial|x|^{\alpha}}-v\frac{\partial f(x,t)}{\partial x}-\wp_{\mathrm{fa}}(t)
\delta(x).
\label{SinkFFPE}
\end{equation}
Here, $v$ denotes the constant external bias, $K_{\alpha}$ is the generalized
diffusion coefficient \cite{report}, and $\wp_{\mathrm{fa}}(t)$ is the density
of first arrival \cite{JPA2003}, as shown below. The fractional derivative of
Riesz-Weyl form is defined in terms of its Fourier transform, $\int_{-\infty}^{
\infty}\exp(ikx)\left[\partial^{\alpha}/\partial|x|^{\alpha}\right]f(x,t)dx=-|k|^{
\alpha}f(k,t)$, where $f(k,t)=\int_{-\infty}^{\infty}\exp(ikx)f(x,t)dx$ is the
Fourier transform of $f(x,t)$ \cite{report}. Eq.~(\ref{SinkFFPE}) is completed
with the initial condition $f(x,0)=\delta(x-x_0)$ of placing the searcher at $x_0$.
If $v$ is positive, the drift is directed towards positive $x$, that is, the
dynamic equation (\ref{SinkFFPE}) describes the situation of Fig.~1.

By rescaling of variables (see App.~\ref{dimensionless}) we obtain the
dimensionless analog of Eq.~(\ref{SinkFFPE}),
\begin{equation}
\frac{\partial\overline{f}(x,t)}{\partial\overline{t}}=\frac{\partial^{\alpha}
\overline{f}(x,t)}{\partial|\overline{x}|^{\alpha}}-\overline{v}\frac{\partial
\overline{f}(x,t)}{\partial\overline{x}}-\overline{p}_{\mathrm{fa}}(\overline{t})
\delta(\overline{x}),
\label{FFPEdimles}
\end{equation}
where $\overline{v}=v\sigma^{\alpha-1}/K_{\alpha}$ and $\overline{f}(\overline{x}, 0)=\delta(\overline{x}-\overline{x}_0)$. The factor $\sigma$ has the dimension of
length and is chosen as the scaling factor of the LF jump length distribution, as
detailed in App.~\ref{dimensionless}. In what follows we use the dimensionless
variables throughout, but for simplicity we omit the overlines. Without loss of generality we assume that $x_0>0$ in the remainder of this work. Integration of Eq.~(\ref{FFPEdimles}) over the position $x$ produces the first arrival density
\begin{equation}
\wp_{\mathrm{fa}}(t)=-\frac{d}{dt}\int_{-\infty}^{\infty}f(x,t)dx.
\end{equation}
Thus, $\wp_{\mathrm{fa}}(t)$ is indeed the negative time derivative of the survival
probability $\int_{-\infty}^{\infty}f(x,t)dx$.

In analogy to the bias-free case \cite{JPA2003} it is straightforward to obtain
the Fourier-Laplace transform of the distribution $f(x,t)$,
\begin{equation}
f(k,s)=\frac{\exp(ikx_0)-\wp_{\mathrm{fa}}(s)}{s+|k|^{\alpha}-ikv}.
\label{fks}
\end{equation}
Here we express the Laplace image $h(s)=\int_0^{\infty}\exp(-st)h(t)dt$ of a
function $h(t)$ by explicit dependence on the Laplace variable $s$. Integration
of Eq.~(\ref{fks}) over the Fourier variable $k$ yields
\begin{eqnarray}
\nonumber
&&\int_{-\infty}^{\infty}f(k,s)dk=f(x=0,s)\\
&&\hspace*{0.8cm}=W(-x_0,s)-W(0,s)\wp_{\mathrm{fa}}(s)=0,
\end{eqnarray}
where $W(x,s)$ is the solution of Eq.~(\ref{SinkFFPE}) without the sink term.
As this expression necessarily equals zero, the first arrival density can be
expressed through
\begin{equation}
\wp_{\mathrm{fa}}(s)=\frac{\int_{-\infty}^{\infty}\exp(ikx_0)\beth dk}
{\int_{-\infty}^{\infty}\beth dk},
\label{pfa}
\end{equation}
where we use the abbreviation
\begin{equation}
\beth\equiv\frac{1}{s+|k|^{\alpha}-ikv}.
\end{equation}
Equation (\ref{pfa}) without bias ($v=0$) was obtained in Ref.~\cite{JPA2003}. An
important observation from Eq.~(\ref{pfa}) is that the first arrival density
vanishes, $\wp_{\mathrm{fa}}(s)=0$, for any $s$ if only $\alpha\leq1$ (for the
proof see App.~\ref{ProofFlat} and \ref{ProofBias}). Thus LF search for
a point-like target will never succeed for $\alpha\leq1$. This property reflects
the transience of LFs with $\alpha<d$, where $d$ is the dimension of the embedding space \cite{sato}.

\subsection{Langevin equation simulations}

For the simulation of LFs we use the Langevin equation approach, which in the discretized version with dimensionless units takes on the form \cite{Slyusarenko}
\begin{equation}
x_{n+1}-x_n=-v\delta t+(\delta t)^{1/\alpha}\xi_{\alpha,1}(n),
\label{LEdiscretized}
\end{equation}
where $x_n$ is the (dimensionless) position of the walker at the $n$-th step, and $\xi_{\alpha,1}(n)$ is a set of random variables with L{\'e}vy stable distribution and the characteristic function
\begin{equation}
p(k)=\int_{-\infty}^{\infty}\exp(ikx)\xi d\xi=\exp\left(-|k|^\alpha\right).
\end{equation}
To obtain a normalized L{\'e}vy stable distribution we employ the standard method detailed in Ref.~\cite{LevyGenerator}.

The modeling of the search process proceeds in the following way. A
walker starts from coordinate $x_0$. Then its position is updated every
step according to the Eq.~(\ref{LEdiscretized}) until it reaches a target
or modeling time exceeds some maximum simulation time limit. Naturally,
the target in simulations can not have size of a point, because then it
will never be found. Hence the target size in simulations should be small
enough in order to get correspondence to results from Eq.~(\ref{pfa}), but
not infinitely small.

\begin{figure}
\includegraphics[width=8cm]{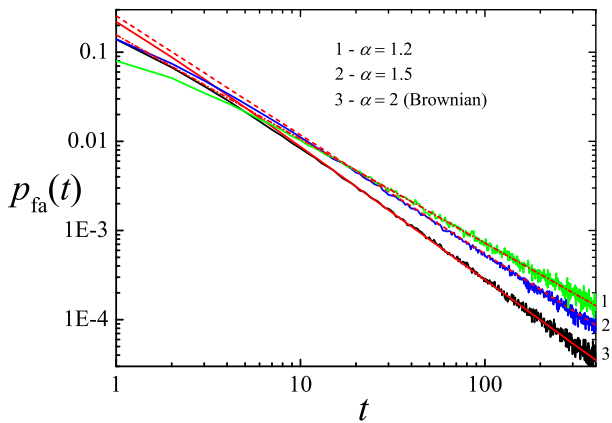}
\caption{First arrival density as function of time for different values of the stable index $\alpha$. The colored curves denote simulations results. The expected asymptotic behavior $\wp_{\mathrm{fa}}(t)\sim t^{-2+1/\alpha}$ is depicted by the red lines. Target sizes were chosen as 0.01 for both $\alpha=2$ and $\alpha=1.5$, and 0.0005 for $\alpha=1.2$.}
\label{PDFfaDiffAlpha}
\end{figure}

\begin{figure}
\includegraphics[width=8 cm]{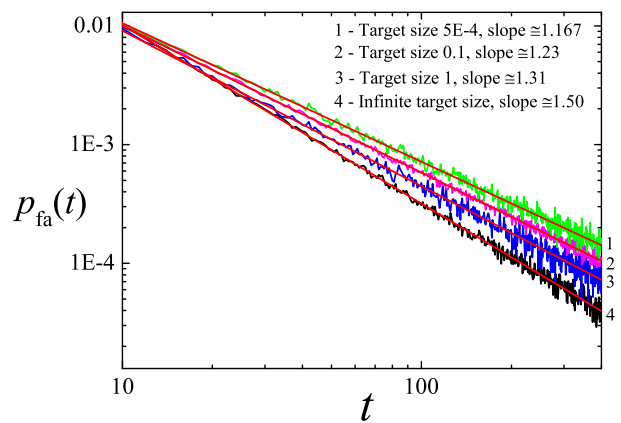}
\caption{For a variety of target sites the colored lines show simulations results for the density of first arrivals, with $\alpha=1.2$ and $x_0=1$. The red lines show fits to the asymptotic power-law form $\wp_{\mathrm{fa}}(t)\simeq t^{-\gamma}$. According to Eq.~(\ref{asymp}) the expected slope for the first arrival is $-1.1\overline{6}$. Thus the smallest target size in the Figure leads to the correct value. Increasing target sizes eventually lead to the universal $-3/2$ Sparre Andersen scaling of the first passage process.}
\label{PDFfaDiffTargets}
\end{figure}

We briefly digress to address an important technical issue. A Brownian walker
always explores the space continuously and therefore localizes any point on the
line. However, in Langevin
equation simulations, we introduce discrete (albeit small) jump lengths
and time steps. Due to this, even for Brownian motion there is always a
non-vanishing probability to overshoot a point-like target. Thus, even for
the Brownian downhill case the simulated value of probability to eventually
find the target becomes less than 1. This effect needs to be remedied by
the appropriate choice of a finite target size. The tradeoff is now that
the target needs to be sufficiently large to avoid the overshoot by the
searcher. At the same time the target should not be too large, otherwise
inconsistencies with our theory based on a point-like target would arise. The
likelihood for leapovers across the target is naturally even more pronounced for
the LF case. As a consistency test for the target size used in the simulations we
check the long time asymptotics
of the first arrival density $\wp_{\mathrm{fa}}(t)$ against the analytical
form given by Eq.~(\ref{asymp}). The results for this test are plotted in
Fig.~\ref{PDFfaDiffAlpha}, showing excellent agreement between the simulations
and the theoretical asymptotic behavior. In Fig.~\ref{PDFfaDiffTargets}
we explicitly show the effect of a varying target size. As the target size
is successively increased, the LF scaling of the first arrival density
for a point-like target, $\wp_{\mathrm{fa}}(t)\sim t^{-2+1/\alpha}$,
is seen to cross over to the universal Sparre-Andersen law for the first
passage of a symmetric random walk process in the semi-infinite domain,
$\wp_{\mathrm{fp}}(t)\sim t^{-3/2}$ \cite{Redner,JPA2003}. We see that it
is possible to choose the target size appropriately such that the results
of the Langevin equation simulations are consistent with the theory.

\section{First arrival and search efficiency in absence of an external bias} 

We first consider the case in absence of the bias $v$ and present the solution
for the first arrival density. Moreover we motivate our choice for the efficiency
used to compare different parameter values for the LF search process.

\subsection{First arrival density and search reliability}

Without external bias Eq.~(\ref{pfa}) can be expressed in terms of the Fox
$H$-function, as detailed in App.~\ref{HfuncV0}. Inverse Laplace transform
(\ref{pfaSolutionFlat}) of the $H$-function allows us to obtain the solution
(\ref{pfatimedomain}) in the time domain. From the latter expression  we get the long time asymptotic behavior of $\wp_{\mathrm{fa}}(t)$,
\begin{equation}
\wp_{\mathrm{fa}}(t)\approx C(\alpha)x_0^{\alpha-1}t^{1/\alpha-2},
\label{asymp}
\end{equation}
where the constant $C(\alpha)$ is given by Eq.~(\ref{Calpha}). In this way we find one of the central results of Ref.~\cite{JPA2003} by using the analytic
approach of the $H$-function formalism.

An important quantity for the following is the search reliability, defined as the cumulative arrival probability
\begin{equation}
\label{reliable}
P=\int_0^{\infty}\wp_{\mathrm{fa}}(t)dt=\wp_{\mathrm{fa}}(s=0).
\end{equation}
It follows from Eq.~(\ref{pfaSolutionFlat}) that without a bias $P=1$, i.e. the searcher will always find the target eventually as long as $\alpha>1$. In other cases it will turn out that $P<1$, that is, the searcher will not always locate the target no matter how long the search process is extended.

In the Brownian case $\alpha=2$, the $H$-function in Eq.~(\ref{pfaSolutionFlat})
according to App.~\ref{BrownDeriv} and \ref{BrownLimit} can be simplified to the well-known result for the first arrival in Laplace domain,
\begin{equation}
\wp_{\mathrm{fa}}(s)=\exp\left(-\sqrt{\frac{sx_0^{2}}{K_2}}\right).
\label{pfaBrown}
\end{equation}
Note that in the Brownian case with finite variance $\langle x^2\rangle$ of relocation lengths, the process of first arrival is identical to that of the first passage \cite{JPA2003}.

\subsection{Search efficiency}

How can one define a good measure for the efficiency of a search process? On a general level, such a definition depends on whether saltatory or cruise foraging is considered \cite{Pitchford}, or whether a single target is present in contrast to a fixed density of targets. For saltatory motion as considered herein a typical definition of the search efficiency is the ratio of the number of visited target sites over the total distance traveled by the searcher \cite{Pitchford},
\begin{equation}
\mathrm{Efficiency}=\frac{\text{visited number of targets}}{\text{average
number of steps}}.
\label{effdef0}
\end{equation}
This definition works well when many targets with a typical inter-target distance are present. The mean number of steps taking in the search process is equivalent to the typical time $\langle t\rangle$ over which the process is averaged. As we here consider the case of a single target, in a first attempt to define the efficiency we could thus reinterpret definition (\ref{effdef0}) as the mean time to reach the target and thus take $\mathrm{Efficiency}=1/\langle t\rangle$, where now $\langle t\rangle$ would correspond to the expectation $\langle t\rangle=\int
_0^{\infty}t\wp_{\mathrm{fa}}(t)dt$. However, in contrast to the situation with a fixed target density, $\langle t\rangle$ diverges for simple Brownian search on a line without bias \cite{Redner}.

For this reason we propose a different measure for the search efficiency, namely
\begin{equation}
\mathcal{E}=\left\langle\frac{1}{t}\right\rangle.
\label{eff}
\end{equation}
Instead of the average search time, we average over the inverse search time. This can be shown to be a useful measure for situations when $\langle t\rangle$ is both finite or diverging. Using the relation
\begin{equation}
\int_{0}^{\infty}\exp(-st)\frac{g(t)}{t}dt=\int_{s}^{\infty}g(u)du
\end{equation}
it is straightforward to show that
\begin{equation}
\mathcal{E}=\int_0^{\infty}\wp_{\mathrm{fa}}(s)ds.
\label{effint}
\end{equation}

As an example, consider the efficiency $\mathcal{E}$ of a Brownian walker without bias. With Eq.~(\ref{pfaBrown}) we find
\begin{equation}
\label{browneff}
\mathcal{E}=\int_0^{\infty}\exp\left(-\sqrt{\frac{s}{K_2}}x_0\right)ds=\frac{2
K_2}{x_0^2},
\end{equation}
where for this equation we restored dimensionality. This is the classical result for a normally diffusive process: increasing diffusivity of the searcher improves the search efficiency per unit time.

Below we demonstrate the robustness of the new characteristic $\mathcal{E}$ for several concrete cases. We mention that the definition $1/\langle t\rangle$ leads to contradictory results for the biased case, as well. This will be shown in the next section.

\subsection{Search optimization}

We now decree that a given search strategy is optimal when the efficiency
$\mathcal{E}$ of the corresponding search process is maximal. In our case of LF search we define the optimal search as the process with the value of the stable index $\alpha$ for fixed initial condition $x_0$ and fixed bias velocity $v$ leads to the highest value of $\mathcal{E}$. As we will see, an optimal search defined by this criterion is not (always) the same as the most reliable process with maximal search reliability $P$.

For LF search without an external drift the density of first arrival is given by Eq.~(\ref{pfaSolutionFlat}). The search efficiency is obtained by integration
it over $s$,
\begin{eqnarray}
\mathcal{E}=\frac{\alpha}{x_0^\alpha}\cos\left(\pi\left[1-\frac{\alpha}{2}\right]
\right)\Gamma(\alpha).
\label{EffLevyFlat}
\end{eqnarray}
Thus the search efficiency decays quadratically with the initial searcher-target
separation $x_0$ and, depending on the value of $\alpha$, may become non-monotonic.
In the Brownian limit $\alpha=2$ the efficiency is $\mathcal{E}=2/x_0^2$,
consistent with the above result (\ref{browneff}). In the Cauchy limit $\alpha\to1$
the efficiency drops to zero.

\begin{figure}
\includegraphics[width=8cm]{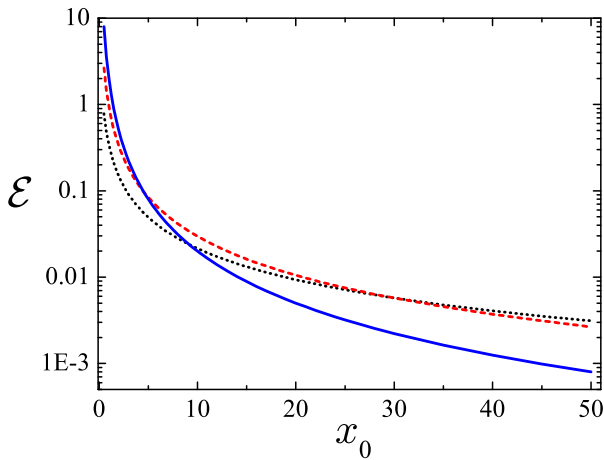}
\caption{L\'evy flight search efficiency as function of the initial position $x_0$, given by Eq.~(\ref{EffLevyFlat}), for three values of $\alpha$: $\alpha=1.2$,
dotted black curve; $\alpha=1.5$, red dashed curve; Brownian case $\alpha=2$, blue continuous curve.}
\label{EffFlatVarX}
\end{figure}

Fig.~\ref{EffFlatVarX} shows the efficiency $\mathcal{E}$ as function of the initial searcher-target distance $x_0$, for fixed values of the power-law
exponent $\alpha$. We observe a strong dependence on $x_0$, the strongest
variation being realized for the Brownian case with $\alpha=2$. For close initial distances ($x_0\lesssim5$) the Brownian strategy is the most efficient process. However, with increasing $x_0$ at first LFs with $\alpha=1.5$ become more efficient than Brownian motion, and for $x_0\gtrsim30$ the strategy with $\alpha=1.2$ outperforms all the others. This behavior is expected as for longer initial separations the occurrence of long jumps increases with decreasing $\alpha$, and thus fewer steps lead the searcher closer to the target. For short initial separations the occurrence of long jumps would lead to leapovers and thus to a less efficient arrival to the target.

\begin{figure}
\includegraphics[width=8cm]{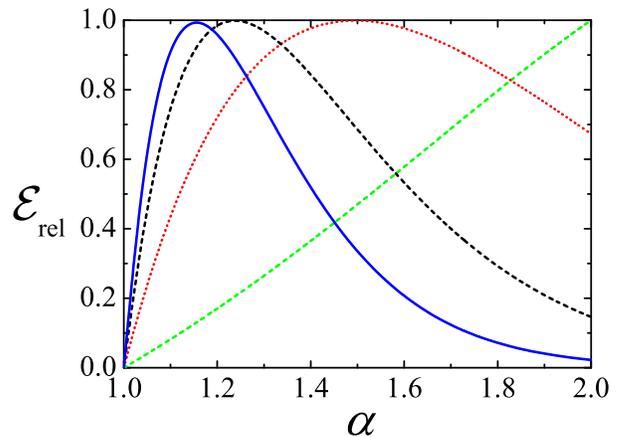}
\caption{Relative efficiency for LF search as a function of the power-law exponent $\alpha$ according to Eq.~(\ref{EffLevyFlat}), displayed for the initial searcher-target separations $x_0=1$ (green dashed curve), $x_0=10$ (red dotted curve), $x_0=100$ (black dashed curve), and $x_0=1000$ (blue continuous curve).}
\label{EffFlatVarAlpha}
\end{figure}

Due to the strong dependence on the initial searcher-target separation $x_0$
the efficiency between different strategies should be compared for a given
value of $x_0$. This will be done in the following. Additional insight can be obtained from the relative efficiency
\begin{equation}
\mathcal{E}_{\mathrm{rel}}=\frac{\mathcal{E}(\alpha)}{\mathcal{E}(\alpha_{
\mathrm{opt}})}
\end{equation}
for a given $x_0$ which is the ratio of the efficiency for some given exponent
$\alpha$ over the maximum efficiency for this initial separation for the
corresponding value $\alpha_{\mathrm{opt}}$. In Fig.~\ref{EffFlatVarAlpha} we show this relative efficiency as function of the stable exponent $\alpha$ of the jump length distribution. The value $\mathcal{E}_{\mathrm{rel}}=1$ is obviously assumed at $\alpha=\alpha_{\mathrm{opt}}$. Fig.~~\ref{EffFlatVarAlpha}
exhibits a very rich behavior. Thus, when the searcher is originally close to the target (here $x_0=1$) the Brownian strategy turns out to be the most efficient, and the functional form of $\mathcal{E}_{\mathrm{rel}}$ is completely monotonic. For growing initial separation, however, the highest efficiency occurs for smaller values of $\alpha$. For instance, the maximum efficiency shifts from $\alpha_{\mathrm{opt}}\approx1.5$ for $x_0=10$ to $\alpha_{\mathrm{opt}}\approx
1.15$ for $x_0=1000$. In particular, for large separations the optimal stable
index approaches the value $\alpha_{\mathrm{opt}}=1$ obtained earlier for
different LF search scenarios 
\cite{viswanathan,PNAS2008,michael,bartumeus1,bartumeus2,reynolds1,reynolds2}.

\begin{figure}
\includegraphics[width=8cm]{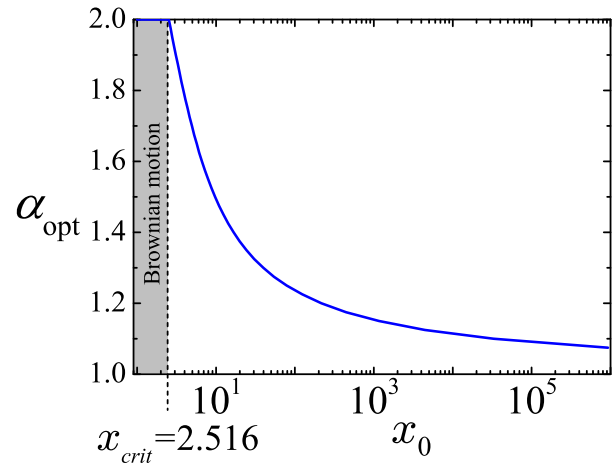}
\caption{Optimal power-law exponent $\alpha_{\mathrm{opt}}$ as a function of the initial searcher-target distance $x_0$, as described by Eq.~(\ref{OptAlpha}).}
\label{EffFlatOptAlpha}
\end{figure}

The second striking observation is that for larger initial separations the
dependence of $\mathcal{E}_{\mathrm{rel}}$ on $\alpha$ is no longer monotonic.
An implicit expression for $\alpha_{\mathrm{opt}}$ is obtained from the relation
\begin{equation}
\frac{d\mathcal{E}(\alpha)}{d\alpha}=0.
\end{equation}   
The result can be phrased in terms of the implicit relation
\begin{equation}
x_0=2\exp\left(\frac{1}{\alpha_{\mathrm{opt}}}+\frac{1}{2}\psi\left(\frac{\alpha_{
\mathrm{opt}}}{2}\right)+\frac{1}{2}\psi\left(\frac{1-\alpha_{\mathrm{opt}}}{2}
\right)\right).
\label{OptAlpha}
\end{equation}
Here $\psi$ denotes the digamma function. From this relation we can use symbolic
mathematical evaluation to obtain the functional behavior of the optimal L{\'e}vy
index $\alpha_{\mathrm{opt}}$ as function of the initial searcher-target distance
$x_0$. The result is shown in Fig.~\ref{EffFlatOptAlpha}. Two distinct phenomena
can be observed: first, the behavior at long initial separations $x_0$ demonstrates
the convergence of the optimal exponent $\alpha_{\mathrm{opt}}$ to the Cauchy value $\alpha_{\mathrm{opt}}=1$. Second, the optimal search is characterized by an increasing value for $\alpha_{\mathrm{opt}}$ when the initial separation shrinks, and we observe a transition at some finite value $x_0$: for initial distances $x_0$ between searcher and target that are smaller than some critical value $x_{\mathrm{\mathrm{crit}}}$, Brownian search characterized by $\alpha_{\mathrm{
opt}}=2$ optimizes the search. In our dimensionless formulation, we deduce from the functional behavior in Fig.~\ref{EffFlatOptAlpha} that $x_{\mathrm{crit}}
\approx2.516$.

\begin{figure*}
\includegraphics[height=6.4cm]{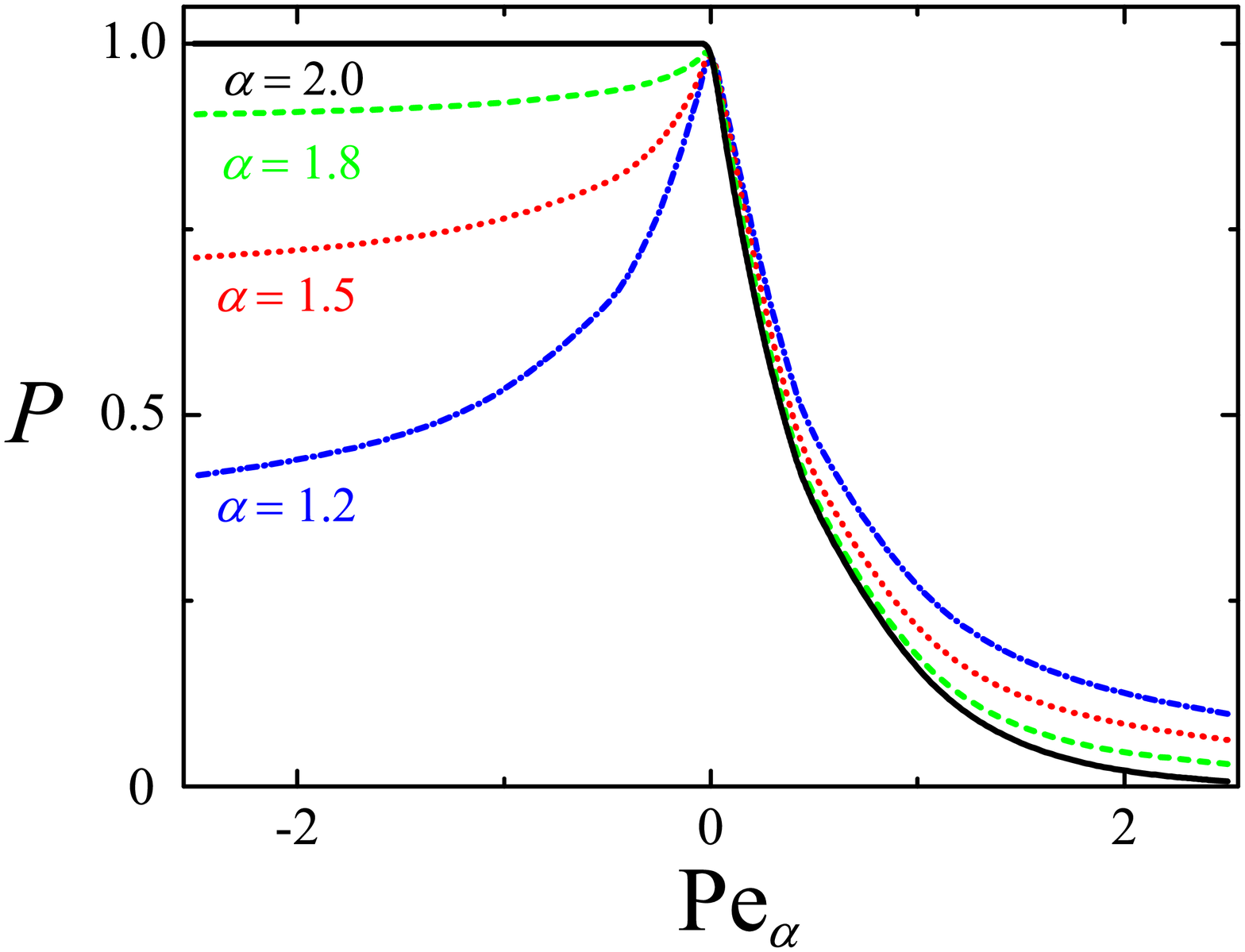}
\hspace*{0.6cm}
\includegraphics[height=6.4cm]{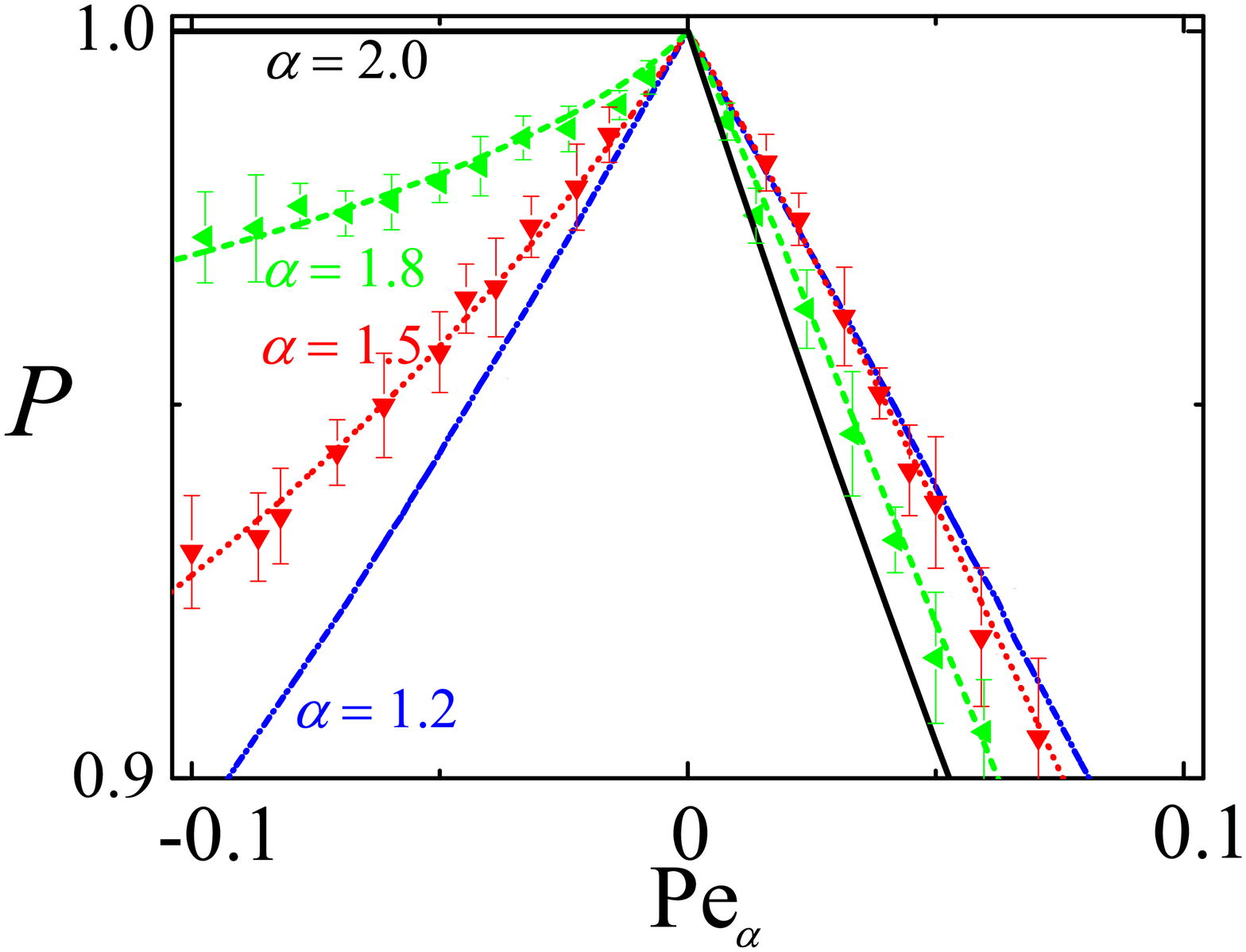}
\caption{Functional dependence of the search reliability $P$ on the generalized
P{\'e}clet number $\mathrm{Pe}_{\alpha}$ defined in Eq.~(\ref{PeDefin}). The lines are obtained numerically from expression (\ref{pfa}), and the symbols in the the zoomed-in region in the right panel are results of Langevin equation
simulations. The lines represent the values $\alpha=2$ (Brownian case, black
continuous line), $\alpha=1.8$ (green dashed line), $\alpha=1.5$ (red dotted line), and $\alpha=1.2$ (blue dashed-dotted line).}
\label{Pe}
\end{figure*}

\section{First arrival and search efficiency in presence of an external bias}

We now consider the case when an external bias initially either pushes the
searcher towards or away from the target, the downhill and uphill scenarios.
In the uphill regime, we can understand that both the Brownian and the LF
searcher may never reach to the target. However, as we will see, due to the
presence of leapovers a LF searcher may also completely miss the target when
we consider the downhill scenario.

\subsection{Search reliability}

We can quantify to what extent a search process will ever locate the target in terms of the search reliability $P$ defined in Eq.~(\ref{reliable}). We obtain this quantity from the first arrival density. We start with the Brownian case for $\alpha=2$, for which the arrival density $\wp_{\mathrm{fa}}$ can be calculated explicitly (see App.~\ref{BrownDeriv} and Ref.~\cite{Redner} for the derivation). In the Laplace domain, it reads
\begin{equation}
\wp_{\mathrm{fa}}(s)=\exp\left(-\frac{x_0v}{2K_2}-x_0\sqrt{\frac{v^2}{4K_2^2}+
\frac{s}{K_2}}\right),
\label{pfabrownbias}
\end{equation}
where we again turned back to dimensional variables to see the explicit dependence
on the diffusivity $K_2$. Thus, in the downhill case with $v<0$ we find that due to the relation $P=p_{fa}(s=0)$ the search reliability will always be unity, $P=1$: in the downhill case the Brownian searcher will always hit the target. In the opposite, uphill case with $v>0$, the result
\begin{equation}
\label{barrier}
P=e^{-vx_0/K_2}
\end{equation}
for the search reliability has the form of a Boltzmann factor ($K_2\propto k_BT$) and exponentially suppresses the location of the target. In this Brownian case we can therefore interpret $P$ as the probability that the thermally driven searcher crosses an activation barrier of height $\propto vx_0$.

For the general case of LFs we obtain from Eq.~(\ref{pfa}) via change of variables the Laplace transform
\begin{equation}
\wp_{\mathrm{fa}}(s)=\frac{\int_{-\infty}^{\infty}\exp(ik)\daleth dk}{\int_{
-\infty}^{\infty}\daleth dk}
\label{pfaPeclet}
\end{equation}
of the first arrival density, where we use the abbreviation
\begin{equation}
\daleth=\frac{1}{sx_0^{\alpha}+|k|^{\alpha}-i2\mathrm{Pe}_{\alpha}k}.
\end{equation}
In expression (\ref{pfaPeclet}) we introduced the generalized P{\'e}clet number for the case of LFs,
\begin{equation}
\mathrm{Pe}_{\alpha}=\frac{1}{2}vx_0^{\alpha-1}.
\label{PeDefin}
\end{equation}
In the Brownian limit $\alpha=2$ and after reinstating dimensional units we recover the standard P{\'e}clet number $\mathrm{Pe}_2=vx_0/(2K_2)$, where the factor two is a matter of choice \cite{Redner}.

\begin{figure*}
\includegraphics[height=6.4cm]{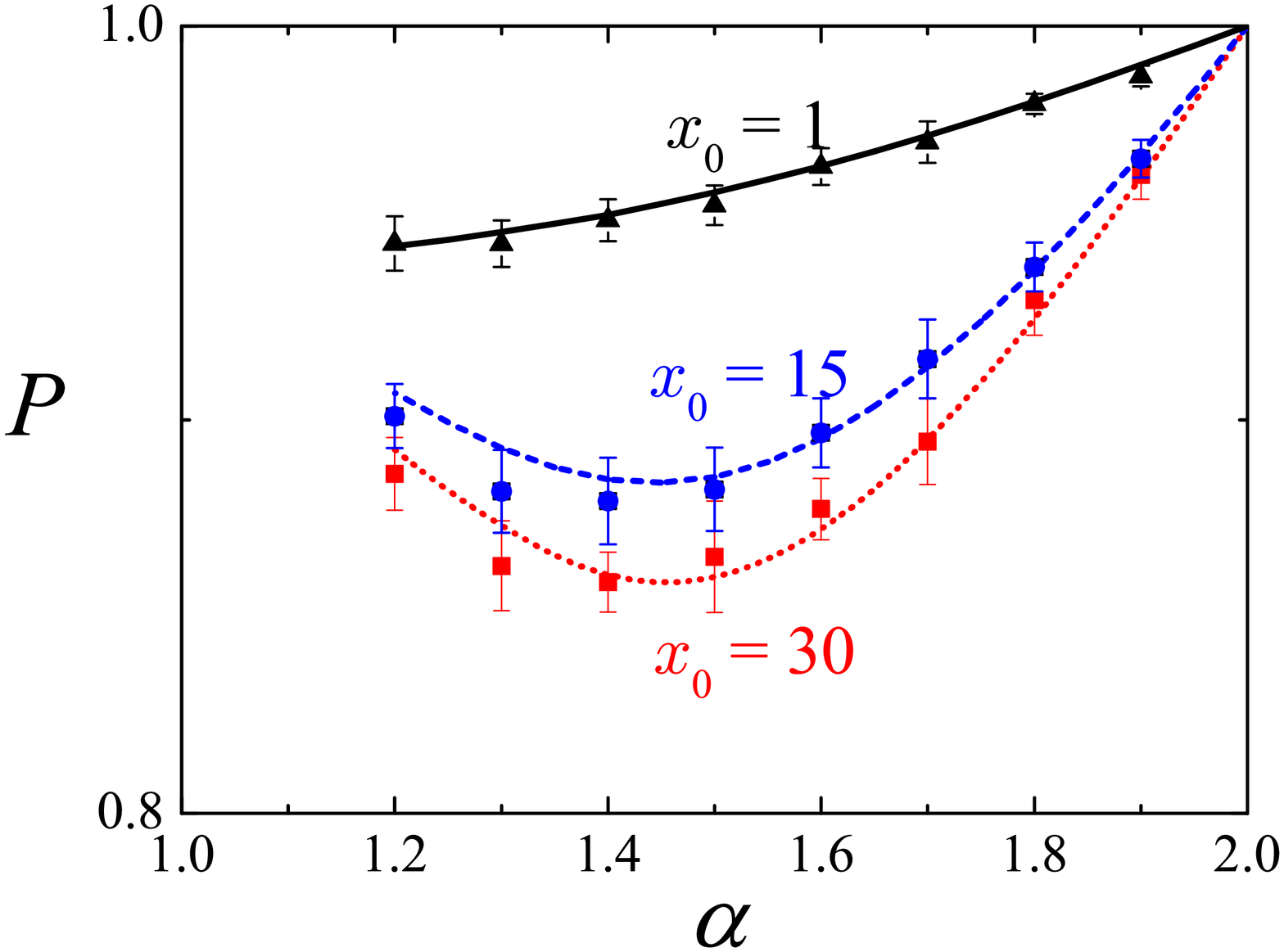}
\includegraphics[height=6.4cm]{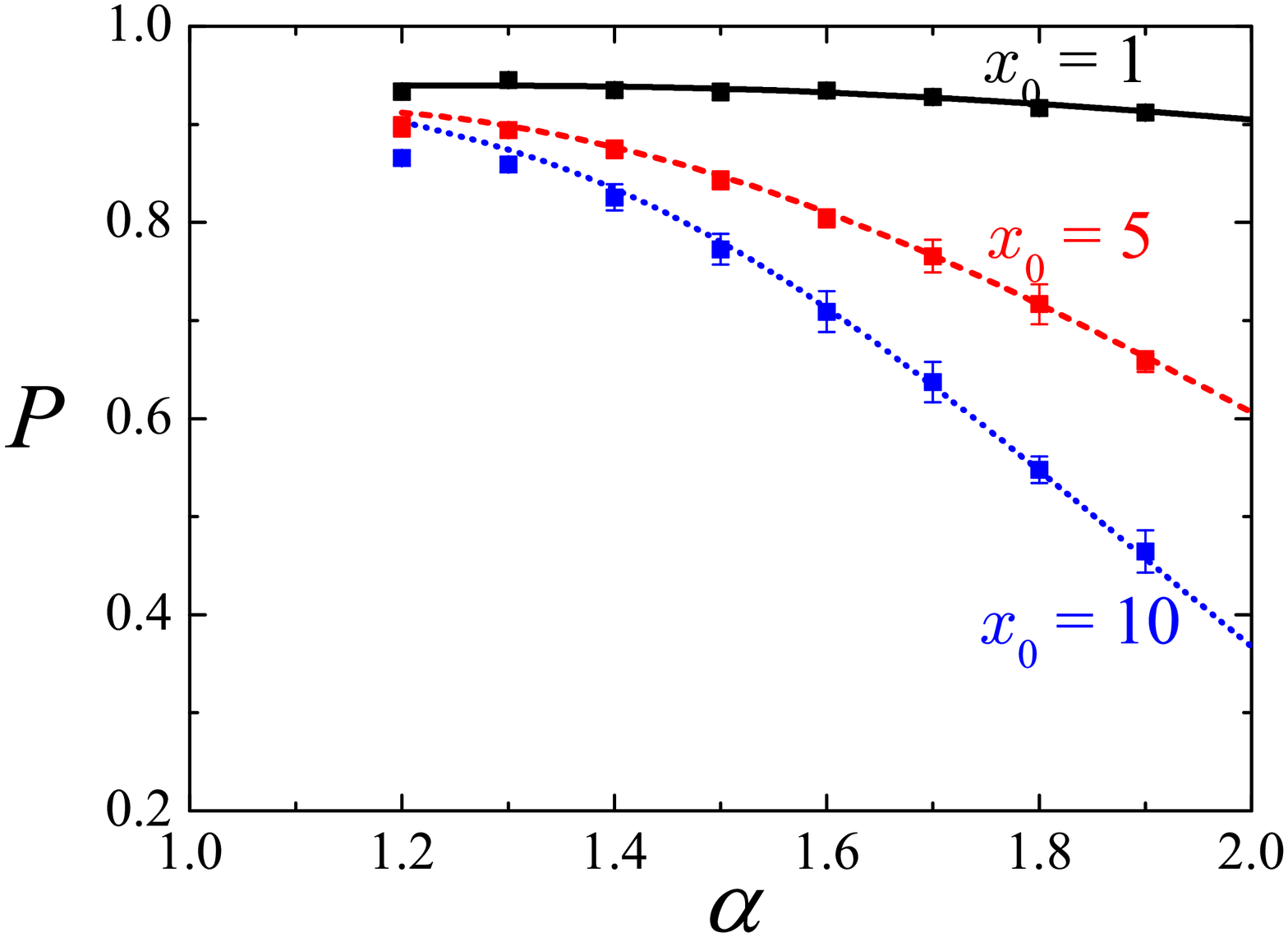}
\caption{Dependence of the search reliability $P$ on the LF power-law exponent
$\alpha$. Left: downhill case, initially the searcher is advected in direction
of the target. Right: uphill case. We show the result for three different
initial conditions. Left: $x_0=30$ (red dotted curve), $x_0=15$ (blue dashed curve), and $x_0=1$ (black continuous curve). Right: $x_0=10$ (blue dashed curve), $x_0=5$ (red dotted curve), $x_0=1$ (black continuous curve). In all cases, $v=-0.1$ (left) and $v=0.1$ (right). The lines are obtained from numerical evaluation of expression (\ref{pfa}), and the symbols denote Langevin equation simulations results.}
\label{x30}
\end{figure*}  

In Fig.~\ref{Pe} we depict the functional behavior of the search reliability $P$ for four different values of the L{\'e}vy index $\alpha$ including the Brownian case $\alpha=2$. The cumulative probability $P$ depends only on the generalized P{\'e}clet number, as can be seen from expression (\ref{pfaPeclet}) when we take the relevant limit $s\to0$. In both panels of Fig.~\ref{Pe} the left semi-axes with negative $\mathrm{Pe}_{\alpha}$ values correspond to the downhill case, in which the searcher is initially advected in direction of the target, while the right semi-axes pertain to the uphill scenario. The continuous lines correspond to the numerical solution of Eq.~(\ref{pfa}), and the symbols represent results based on Langevin equation simulations. In these simulations the values of the search reliability $P$ were obtained as a ratio of the number of searchers that eventually located the target over the overall number of the released 10,000 searchers. To estimate the error of the simulated value for the search reliability, we calculated
$P$ for each consecutive 1000 runs and then determined the standard deviation of the mean value of these 10 results.

According to Fig.~\ref{Pe} for the case of uphill search the search reliability
is worst for the Brownian walker and improves continuously for decreasing value of the stable index $\alpha$. This is due to the activation barrier (\ref{barrier}) faced by the Brownian walker. For LFs this barrier is effectively reduced due to the propensity for long jumps. The reduction of the resulting jump length $x-v\tau$, where $\tau$ is the typical duration of a single jump, becomes more and more insignificant for increasing jump lengths $x$. This is why the efficiency continues to improve until the Cauchy case is reached. Quantitatively, however, we realize that for increasing generalized P{\'e}clet number even for LFs the value of the search reliability quickly decreases to tiny values, and that the absolute difference between the different search strategies is not overly significant.

In the downhill case Fig.~\ref{Pe} demonstrates that the Brownian searcher will always locate the target successfully and thus return $P=1$ in agreement with previous findings \cite{Redner}. In contrast, the search reliability decreases clearly with growing magnitude $|\mathrm{Pe}_{\alpha}|$. This decrease worsens with decreasing stable exponent $\alpha$. The reason for this is the growing tendency for leapovers of LFs with decreasing $\alpha$. Once the LF searcher overshoots the target, it is likely to drift away quickly from the target and never return to its neighborhood. Overall, the functional dependence of $P$ on the  generalized P{\'e}clet number $\mathrm{Pe}_{\alpha}$  becomes non-trivial once $\alpha<2$. From Fig.~\ref{Pe} we conclude that if the main criterion for the search is the eventual location of the target, that is, a maximum value of the search reliability $P$, without prior knowledge the gain for a Brownian searcher in the downhill case is higher than the loss in the opposite case: if we do not know the relative initial position to the target, the Brownian search algorithm will on average be more successful.

In Fig.~\ref{x30} we now turn to the dependence of the search efficiency $P$ on the L{\'e}vy index $\alpha$ for fixed magnitude of the external bias $v$. In each case we display the behavior for three different values of the initial distance $x_0$ between searcher and target. In the downhill scenario we observe a remarkable non-monotonic behavior for larger value for $x_0$. Namely, the search efficiency drops when $\alpha$ gets smaller than the Brownian value $\alpha=2$, for which $P=1$. While for the small initial separation $x_0=1$ this drop is continuous, for the larger values of $x_0$ this trend is turned around, and the search efficiency grows again. Due to the extremely slow convergence of both the Langevin equation simulations and the numerical evaluation of Eq.~(\ref{pfa}) despite all efforts we were not able to infer the continuation of the $P$-curve for $\alpha$-values smaller than $1.2$ and thus, in particular, what the limiting value at the Cauchy case $\alpha=1$ is. What could be the reason for this non-monotonicity in the $P$ versus $\alpha$ dependence? Similar to the existence of an optimal $\alpha$-value intermediate between the Brownian and Cauchy cases $\alpha=2$ and $\alpha=1$, respectively, for the search reliability we here find a worst-case value for $\alpha$. This value represents a negative tradeoff of the target overshoot property and insufficient propensity to produce sufficiently long jumps to recover an accumulated activation barrier from a downstream location
as seen from the target. In the uphill case the dependence is monotonic: here long jumps become helpful to overcome the activation towards the target. Thus the search efficiency increases when $\alpha$ becomes smaller and approaches the Cauchy value $\alpha=1$. The value for $P$ significantly drops with increasing
value of the initial searcher-target separation $x_0$. Consistently with the previous observations on $P$ the Brownian case fares worst and leads to the smallest value of $P$.

\subsection{Search efficiency}

For a Brownian searcher in the presence of an external bias, the mean search time $\langle t\rangle$ can be computed via
\begin{equation}
\langle t\rangle=-\left.\frac{\partial\wp_{\mathrm{fa}}(s)}{\partial s}\right|_{
s=0}.
\end{equation}
For the downhill case this value is given by the classical result
\begin{equation}
\langle t\rangle= \frac{x_0}{|v|}.
\end{equation}
The diffusing searcher moves towards the target as if it were a classical particle, the search time being given as the ratio of the distance over the (drift) velocity. It is independent of the value of the diffusion constant. For the uphill case ($v>0$), we find the known result
\begin{equation}
\langle t\rangle=\frac{x_0}{v}\exp\left(-\frac{vx_0}{K_2}\right).
\end{equation}
The latter value is in fact smaller than the one for the downhill case. How can this be? The explanation of this seeming paradox comes from the qualitative
difference in the nature of these averages. In the first scenario the search
reliability is unity, that is, the walker always arrives at the target. In the uphill case only successful walkers count, that is, the average is conditional.
This explains the seeming contradiction with common sense \cite{Tikhonov}. As we can see from this discussion is that the ready choice $1/\left<t\right>$ as a measure for the search efficiency would state that the uphill motion is more efficient than the downhill one. This definition would obviously not make much sense. We show that our definition of the search efficiency, Eq.~(\ref{eff}), is a reasonable measure in this case. With the use of Eqs.~(\ref{effint}) and  (\ref{pfabrownbias}) we find
\begin{equation}
\label{EffBrow}
\mathcal{E}=\frac{2K_2}{x_0^2}\left(1+\frac{|v|x_0}{2K_2}\right)\left\{
\begin{array}{ll}1,& v\leq0\\[0.2cm]
\exp\left(-vx_0/K_2\right),& v\geq0\end{array}\right..
\label{EffBrowV}
\end{equation}
Indeed, we see that our expression for the efficiency shows that the downhill
motion is more efficient than going uphill for the same initial separation
$x_0$.

\begin{figure}
\includegraphics[width=8.6 cm]{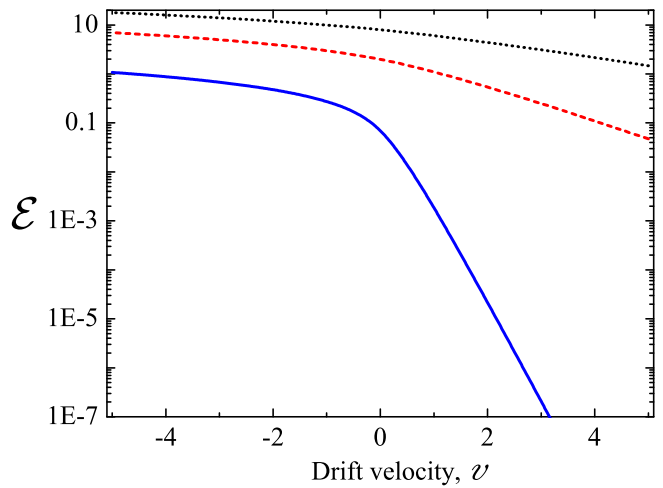}
\caption{Search efficiency as function of the drift velocity $v$ from numerical
evaluation of Eq.~(\ref{EffBrowV}), for three values of the initial searcher-target
separation $x_0$: $x_0=0.5$ (dotted black curve), $x_0=1$ (red dashed curve), and $x_0=5$ (blue continuous curve).}
\label{EffBiasBrowX}
\end{figure} 

\begin{figure}
\includegraphics[width=8.6 cm]{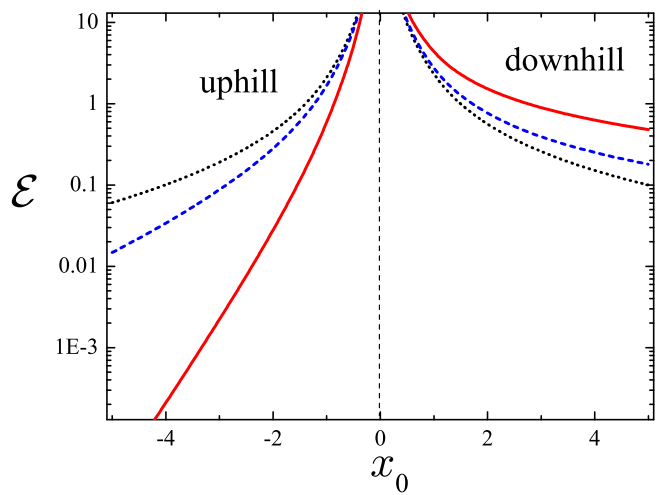}
\caption{Search efficiency as function of the initial position $x_0$ calculated
from Eq.~(\ref{EffBrowV}) for three values of the drift velocity $v$: $v=-0.1$
(dotted black curve), $v=-0.5$ (blue dashed curve), and $v=-2$ (red continuous
curve).}
\label{EffBiasBrowV}
\end{figure} 

In Figs.~\ref{EffBiasBrowX} and \ref{EffBiasBrowV} the efficiency $\mathcal{E}$ is plotted for different values of the drift velocity $v$ and the initial separation $x_0$ of searcher and target, respectively. As expected, the increase of the downhill velocity leads to an efficiency growth, and vice versa for the opposite case. By magnitude of the $v$-dependence, the decrease in the search efficiency for the uphill case is much more pronounced than the increase in efficiency for the downhill case. Hence, the dependence on the initial distance $x_0$ becomes increasingly asymmetric.

\subsection{Weak bias expansion for L{\'e}vy flight search}

\begin{figure}
\includegraphics[width=8.6 cm]{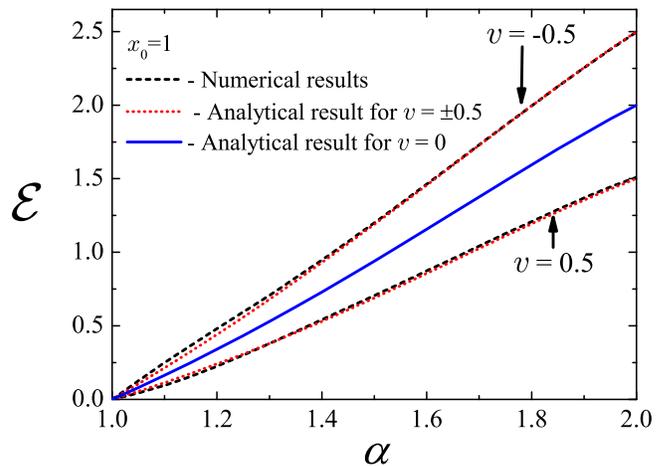}
\caption{Comparison of the search efficiency from the approximate expression
(\ref{EffLevyBias}) and from numerical integration of expression (\ref{pfa})
over $s$, for the initial searcher-target separation $x_0=1$. Results are shown for the cases of zero drift as well as uphill and downhill drift.}
\label{ApproxComp1}
\end{figure}

In the limit of a weak external bias we can obtain analytical approximations for the search efficiency. Namely, for sufficiently small values of the generalized
P{\'e}clet number $\mathrm{Pe}_{\alpha}$ and nonzero values of the Laplace
variable $s$ the denominators in both integrals of Eq.~(\ref{pfaPeclet}) can be expanded into series. The first order expansion reads
\begin{equation}
\wp_{fa}(s)\simeq\frac{\int_{-\infty}^{\infty}\cos(k)\aleph dk-\int_{-\infty}^{
\infty}2\mathrm{Pe}_{\alpha}k\sin(k)\aleph^2dk}{\int_{-\infty}^{\infty}
\aleph dk},
\label{pfaexp}
\end{equation}
where we define
\begin{equation}
\label{aleph}
\aleph=\frac{1}{sx_0^{\alpha}+|k|^{\alpha}}
\end{equation}
The integrals appearing in expression (\ref{pfaexp}) can be computed by use of the Fox $H$-function technique, as detailed in App.~\ref{AppGenExp}. From the result (\ref{pfaExpansionGen}) we obtain the following expression for the search efficiency,
\begin{equation}
\mathcal{E}=\frac{\alpha}{x_0^\alpha}\left[\cos\left(\pi\left(1-\frac{\alpha}{2}
\right)\right)\Gamma(\alpha)-2\left(1-\frac{1}{\alpha}\right)\mathrm{Pe}_{\alpha}
\right],
\label{EffLevyBias}
\end{equation}
where the first term in the square brackets corresponds to the result for the case without drift ($\mathrm{Pe}_{\alpha}=0$), Eq.~(\ref{EffLevyFlat}). When $\alpha=2$, the Brownian behavior in the small bias limit is recovered, namely, $\mathcal{E}=2/x_0^2(1-\mathrm{Pe}_2)$, consistent with the small $\mathrm{Pe}_2$ expansion of Eq.~(\ref{EffBrowV})). From Eq.~(\ref{pfaExpansionGen}) it follows that for the Brownian case $\alpha=2$, the first arrival density in the Laplace domain has the approximate form
\begin{equation}
\wp_{\mathrm{fa}}(s)\simeq\left(1-\mathrm{Pe}_2\right)\exp\left(-\sqrt{\frac{sx_0
^2}{K_2}}\right),
\end{equation}
which is valid for $s>v^2/K_2$, i.e., for short times. After transforming back to the time domain, we find
\begin{equation}
\wp_{\mathrm{fa}}(t)\simeq\frac{(1-\mathrm{Pe}_2)x_0}{\sqrt{4\pi K_2t^3}}\exp\left(
-\frac{x^2_0}{4K_2t}\right),
\end{equation}
as shown in App.~\ref{BrownDeriv}. This result corresponds to the short time and small P{\'e}clet number limit of the general expression for $\wp_{\mathrm{fa}}(t)$ reported by Redner \cite{Redner}. Thus our expansion (\ref{pfaExpansionGen}) works only at short times. However, the approximate expression (\ref{EffLevyBias}) for the search efficiency itself turns out to work remarkably well, as shown in Fig.~\ref{ApproxComp1}. Here, the behavior described by Eq.~(\ref{EffLevyBias}) is compared with results of direct numerical integration of Eq.~(\ref{pfa}) over $s$. We see an almost exact match for an initial searcher-target separation $x_0=1$. Instead, for $x_0=10$ the agreement becomes worse (not shown here). The explanation is due to the fact that for small initial separations short search times dominate the arrival statistic, while for $x_0$ the arrival is shifted to longer times, and the approximation underlying Eq.~(\ref{EffLevyBias}) does no longer work well. 

\begin{figure*}
\includegraphics[height=6.4cm]{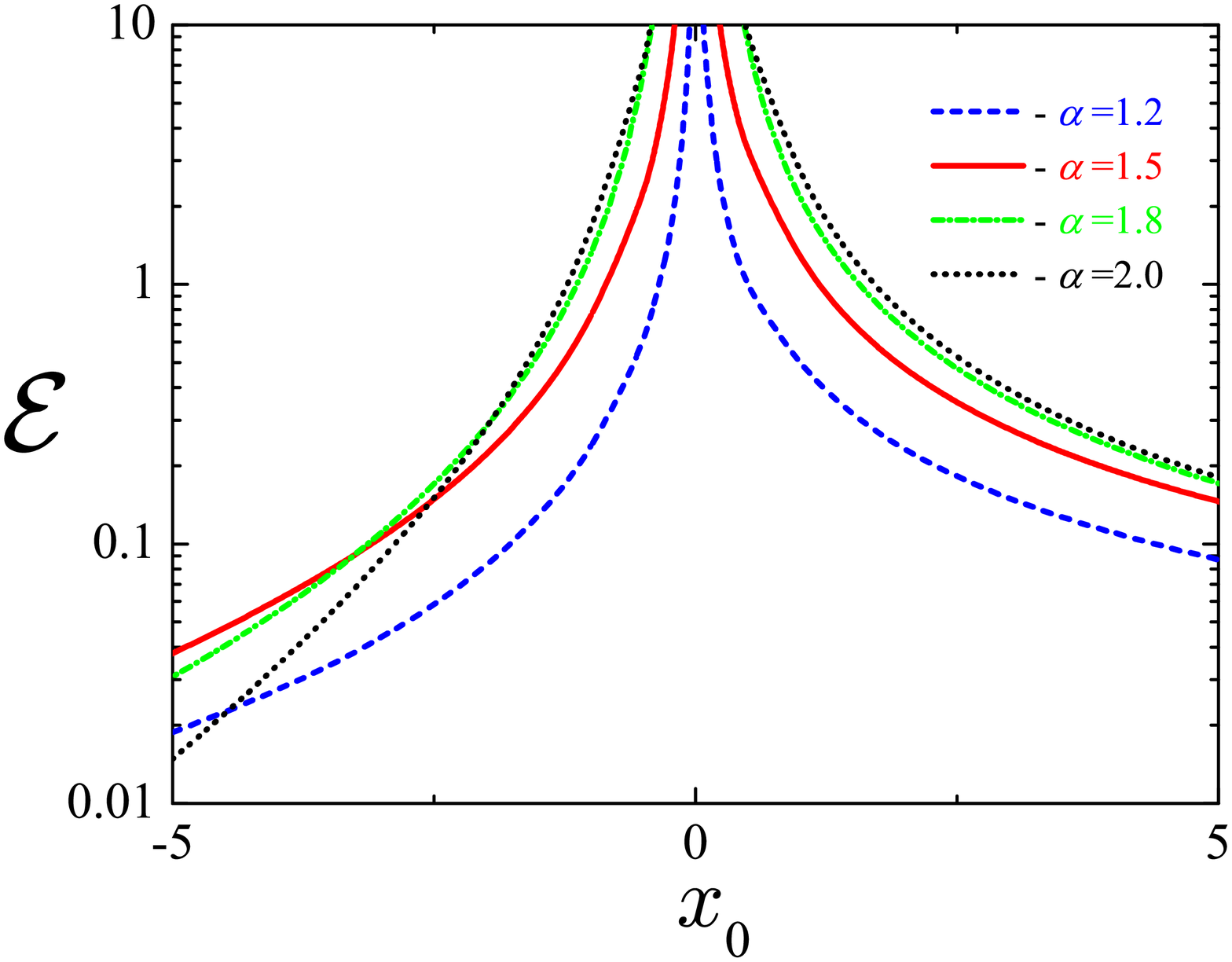}
\hspace{0.6cm}
\includegraphics[height=6.4cm]{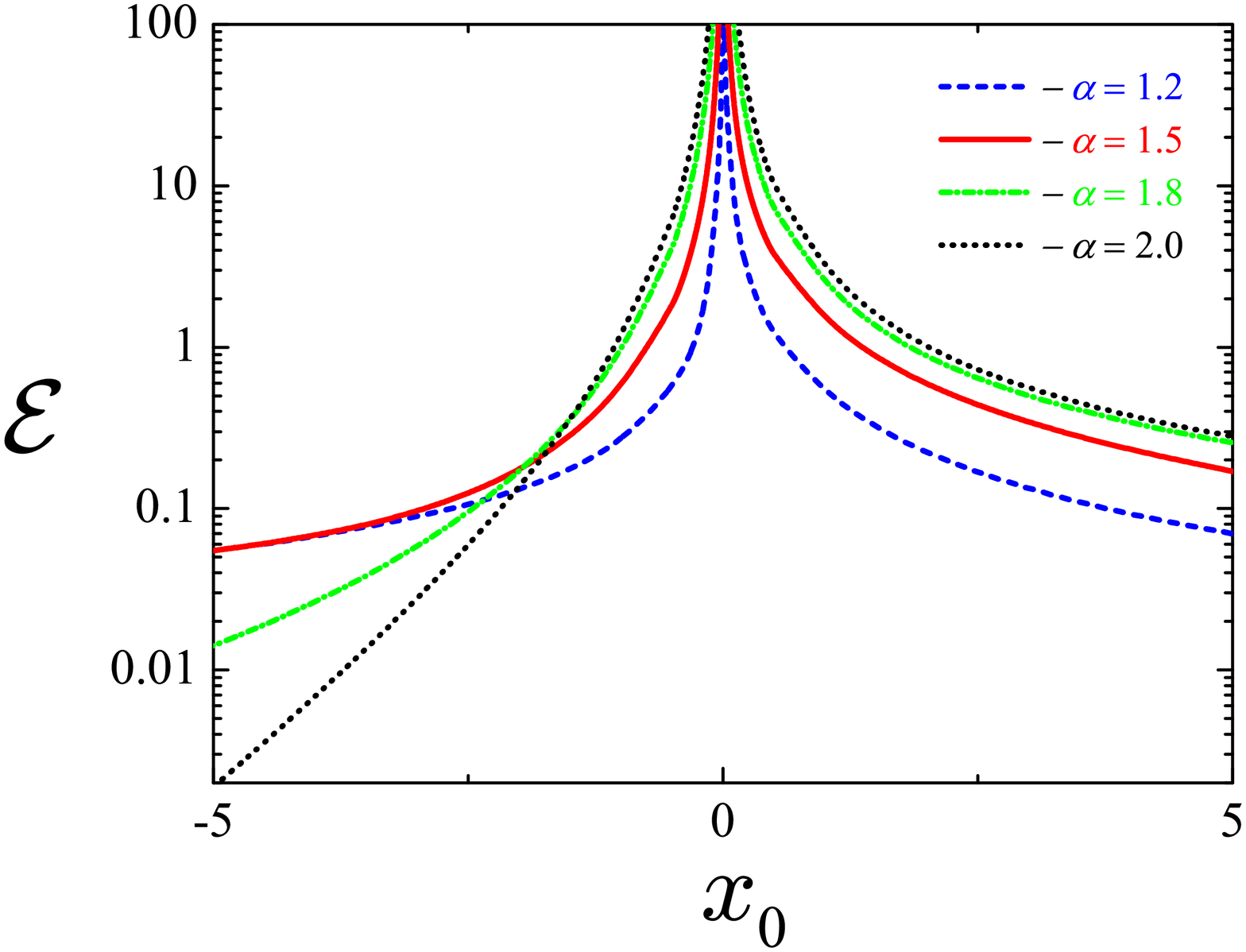}
\caption{Search efficiency in the presence of a bias as function of the initial
searcher-target separation $x_0$ for different values of the power-law exponent:
$\alpha=2$ (green dotted line), $\alpha=1.8$ (red dashed line), $\alpha=1.5$
(black dashed-dotted line), and $\alpha=1.2$ (blue continuous line). Left:
$v=-0.5$. Right: $v=-1$.}
\label{EffBiasLevy1}
\end{figure*}

The presence of an external bias substantially changes the functional form of the efficiency as compared to the unbiased situation. Fig.~\ref{EffBiasLevy1}
shows the search efficiency as function of the initial position for two
different drift velocities and for a variety of values of the power-law exponent
$\alpha$. As expected from what we said before, the dependencies of the search
efficiency with respect to positive and negative initial separations is asymmetric,
as this corresponds to the difference between uphill and downhill cases elaborated
above. Increasing magnitude of the bias effects a more pronounced asymmetry between $x_0$ values with the same absolute value $|x_0|$. For the downhill case the advantage of the Brownian search over LF search persists for all values of $x_0$. As expected, the efficiency drops, however, the general behavior is similar for all $\alpha$ values. For the uphill case we observe a remarkable crossing of the curves. For small initial separations $x_0$ the Brownian search efficiency is highest. Here the activation barrier is sufficiently small such that the continuous Brownian searcher without leapovers locates the target most efficiently. When $x_0$ goes to increasingly negative values, successively LFs with smaller $\alpha$ values become more efficient. In terms of the efficiency we see that for sufficiently large barriers, that is, when the target is initially separated by a considerable uphill distance from the searcher, LFs with smaller $\alpha$ fare dramatically better than processes with larger $\alpha$.

\begin{figure*}
\includegraphics[height=6.4cm]{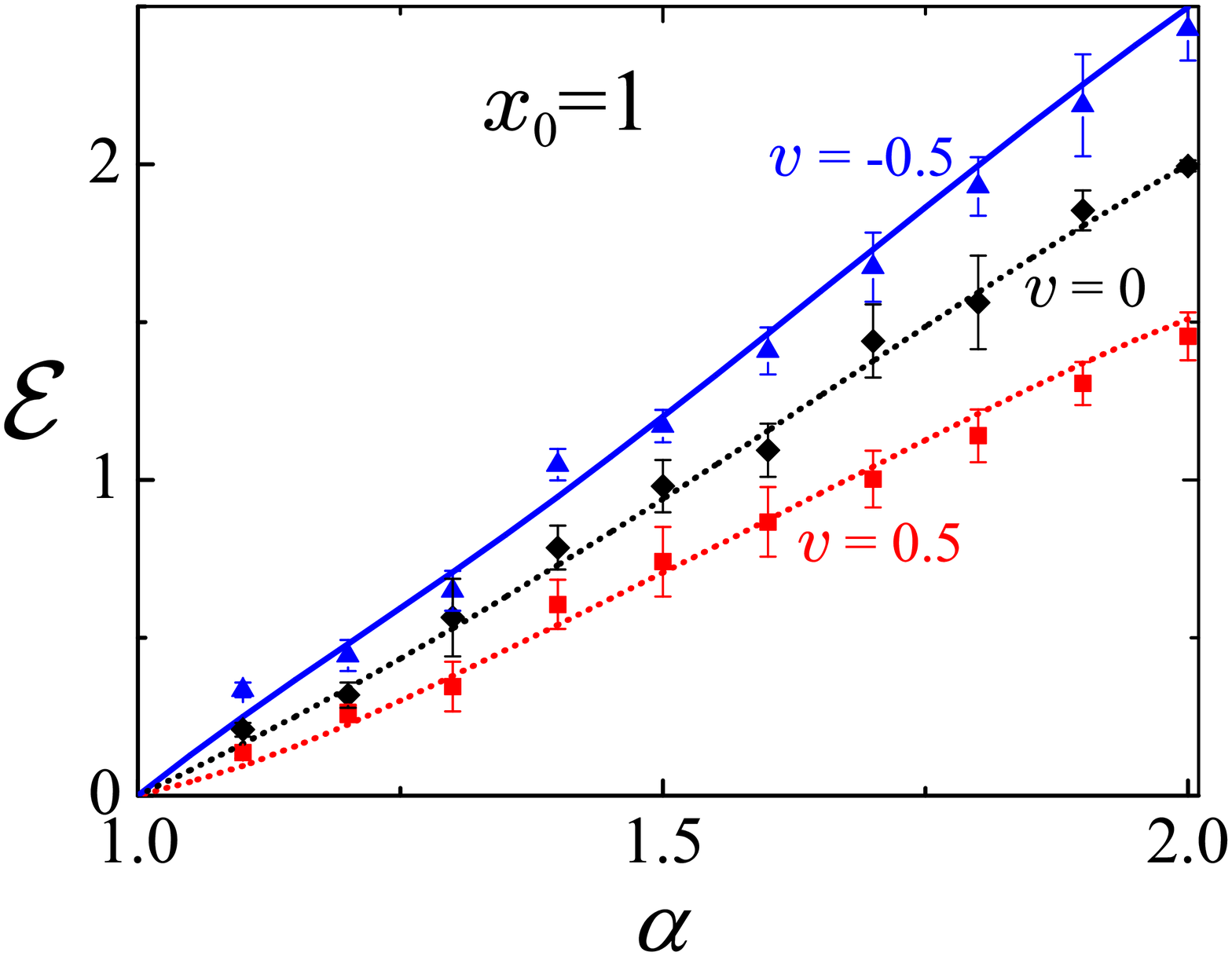}
\hspace{0.6cm}
\includegraphics[height=6.4cm]{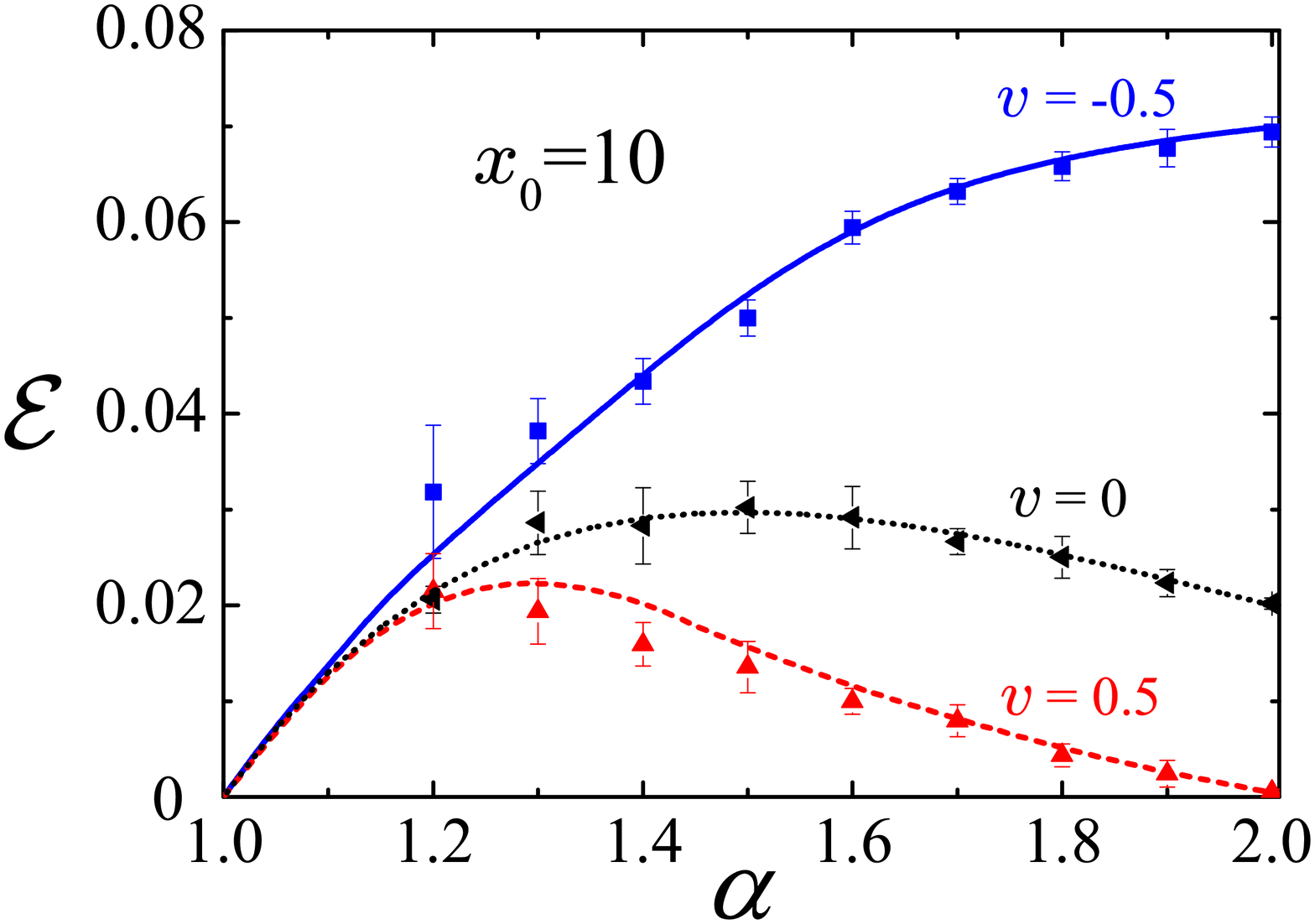}
\caption{Search efficiency as function of the power-law exponent $\alpha$ for initial searcher-target separation $x_0=1$ (left) and $x_0=10$ (right). We show the dependence for the following bias velocities: $v=-0.5$ (blue upper curve), $v=0$ (black center curve), and $v=-0.5$ (red lower curve). Symbols correspond to Langevin equation simulations.}
\label{EffLevyVarAlphaS}
\end{figure*} 

\begin{figure}
\includegraphics[height=6.4cm]{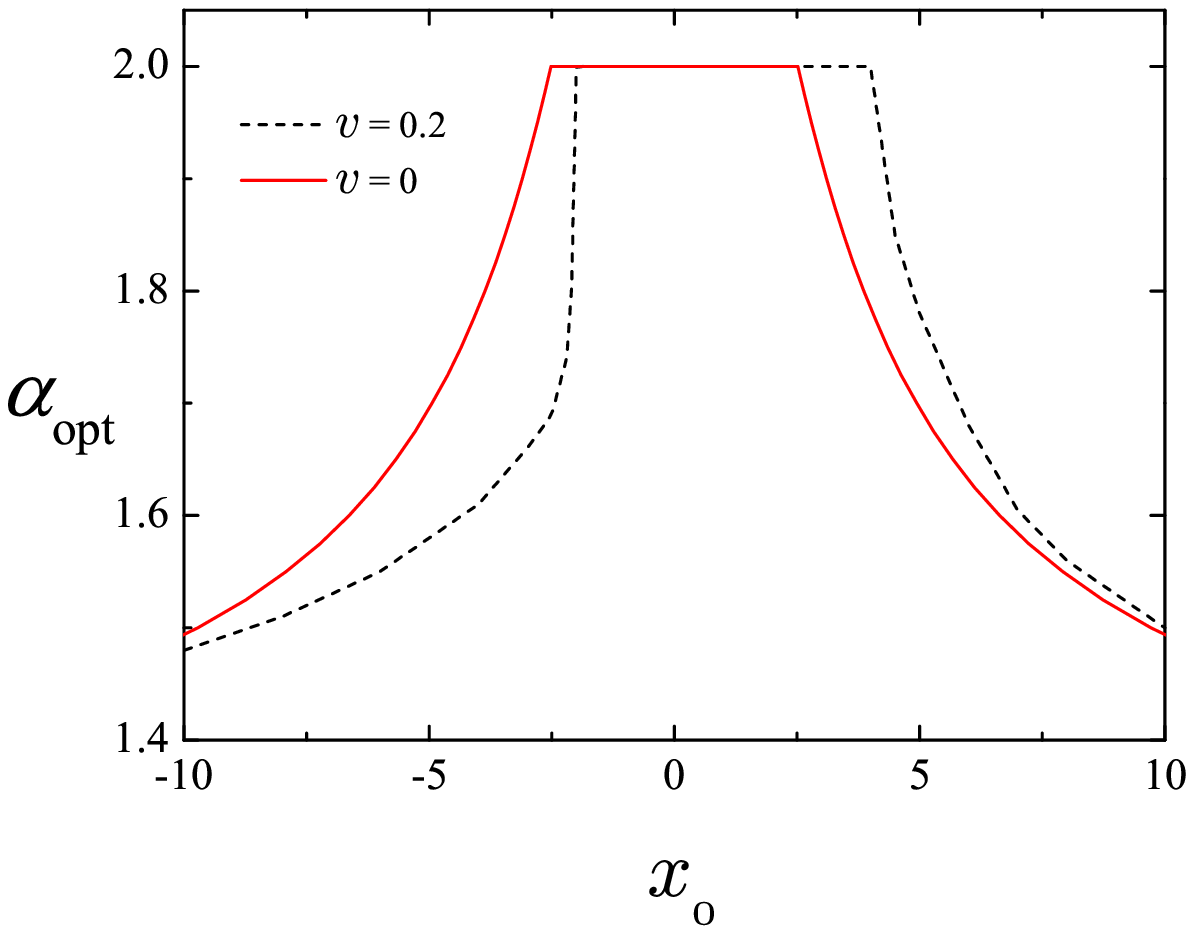}
\caption{Optimal power-law exponent $\alpha_{\mathrm{opt}}$ as a function of
initial searcher-target distance $x_0$ with bias velocity $v=-0.2$ (black
dashed line) and without bias $v=0$ (red continuous line).}
\label{newplot}
\end{figure} 

We further illustrate the behavior of the search efficiency by studying its
functional dependence on the stable index $\alpha$ for different initial distances $x_0$ between searcher and target as well as for different drift velocities in Fig.~\ref{EffLevyVarAlphaS}. Thus, for short initial separation $x_0$ shown in Fig.~\ref{EffLevyVarAlphaS} on the left, the Brownian searcher is always the most efficient for all cases: unbiased, downhill, and uphill. For larger $x_0$ as shown in the right panel of Fig.~\ref{EffLevyVarAlphaS}, the situation changes: in the downhill case the Brownian searcher still fares best. However, already in the unbiased case the LF searchers produce a higher efficiency. An interesting fact is the non-monotonicity of the behavior of the search efficiency, leading to an optimal value for the stable index $\alpha$, whose value depends on the strength of the bias (and the initial separation $x_0$). This $\alpha_{\mathrm{opt}}$ is shifting towards the Cauchy value $\alpha=1$ for increasing uphill bias.

We can also find a qualitative argument for the optimal value $\alpha_{\mathrm{
opt}}$ of the power-law index in the small bias limit. If we denote $\mathcal{E}_0=
\mathcal{E}(x_0,v=0)$, then $\mathcal{E}(x_0,v)\cong\mathcal{E}_0-(\alpha-1)v/x_0$.
In comparison to the unbiased case, that is, the efficiency is reduced in the uphill case and increased in the downhill case, as it should be. Moreover, the correction due to the bias is more pronounced for larger values of $\alpha$. Hence the optimal $\alpha$ necessarily shifts to larger values for the downhill case in comparison with the unbiased situation, and vice versa in the uphill case. 

This can be perfectly illustrated with Fig. \ref{newplot}. The plot shows that application of a bias apparently breaks a symmetry in terms of initial position of a searcher: optimal $\alpha$ values increase for downhill side and drop for uphill side. Thus, the range of $x_0$ values where the Brownian motion is optimal is effectively shifted.

\subsection{Implicit formula for the first arrival density}

We briefly mention a different way to approach the first arrival problem in terms of an implicit expression for the corresponding density $\wp_{\mathrm{fa}}(t)$.
From Eq.~(\ref{pfa}), by inverse Laplace transform we find
\begin{eqnarray}
\nonumber
\int_0^t\wp_{\mathrm{fa}}(t-\tau)d\tau\int_{-\infty}^{\infty}e^{ikv\tau-|k|^{
\alpha}\tau}dk\\
=\int_{-\infty}^{\infty}e^{ikx_0+ikv\tau-|k|^{\alpha}\tau}dk.
\label{alter}
\end{eqnarray}
With the functions $H_i(t)$ defined in App.~\ref{implicitApp}, we rewrite
this relation in the form
\begin{equation}
\int_0^t\wp_{\mathrm{fa}}(t-\tau)H_1(\tau)d\tau=H_2(t),
\end{equation}
such that we arrive at the simple form
\begin{equation}
\wp_{\mathrm{fa}}(s)=\frac{H_2(s)}{H_1(s)}
\label{ratio}
\end{equation}
in terms of the Laplace transforms $H_i(s)$. This is a familiar form for the first passage density for continuous processes \cite{Redner}, and is also known for the first arrival of LFs \cite{JPA2003}. For numerical evaluation or small bias expansions this expression turns out to be useful.

\section{Discussion}

We generalized the prominent L{\'e}vy flight model for the random search of
a target to the case of an external bias. This bias could represent a choice
of the searcher due to some prior experience, a bias in an algorithmic search
space, or simply an underwater current or airflow. To compare the efficiency
of this biased LF search for different initial searcher-target separations
and values of the external bias, we introduced the search efficiency in terms
of the mean of the inverse search time, $\left<1/t\right>$. We confirmed that
this measure is meaningful and in fact more consistent than the traditional
definition in terms of the inverse mean search time, $1/\langle t\rangle$. As a
second measure for the quality of the search process we introduced the search
reliability, the cumulative arrival probability. When this measure is unity,
the searcher will ultimately always locate the target. When it is smaller
than unity, the searcher has a finite chance to miss the target. As shown
here, high search reliability does not always coincide with a high search
efficiency. Depending on what we expect from a search process, either measure
may be more relevant.

In terms of the efficiency we saw that even in absence of a bias the optimal
 strategy crucially depends on the initial separation $x_0$ between the
searcher and the target. For small $x_0$ the Brownian searcher is more
efficient, as it cannot overshoot the target. With increasing $x_0$, however,
the LF searcher needs a smaller number of steps to locate the target and thus
becomes more efficient. In the presence of a bias there is a strong asymmetry
depending on the direction of the bias with respect to the initial location of
the searcher and the target. For the downhill scenario the Brownian searcher
always fares better, as it is advected straight to the target while the LF
searcher may dramatically overshoot the target in a leapover event and then
needs to makes its way back to the target, against the bias. The observed
behaviors can be non-monotonic, leading to an optimal value for the power-law
exponent $\alpha$. For strong uphill bias and large initial separation the
optimal $\alpha$ value is unity, for short separations and downhill scenarios
the Brownian limit $\alpha=2$ is best. There exist optimal values for $\alpha$
in the entire interval $(1,2)$, depending on the exact parameters.

The search reliability for a given value of $\alpha$ solely depends on
the generalized P{\'e}clet number. For unbiased search the searcher
will always eventually locate the target, that is, the search efficiency
attains the value of unity. In the presence of a bias, unity is returned
for the search reliability for a Brownian searcher in the downhill case. It
decays exponentially for the uphill case. For LF searchers with $\alpha$
smaller than two, the probability of leapovers reduces the value of the
search reliability in the downhill case. In the opposite, uphill scenario
the search reliability is larger for LF searchers compared to the Brownian
searcher. The absolute gain in this case, however, was found to be smaller
than the loss to a Brownian competitor in the downhill case. Without prior
knowledge of the bias a Brownian search strategy may turn out to be overall
advantageous. We also found a non-monotonicity of the search reliability as
function of the initial searcher-target separation $x_0$. It
will be interesting to see whether our results for both the search efficiency
and reliability under an external bias turn out similarly for periodic boundary
conditions relevant for finite target densities.

We note here that we analyzed LF search in one spatial dimension. What would be
expected if the search space has more dimensions? For regular Brownian motion
we know that it remains recurrent in two dimensions, that is, the sample
path is space-filling in both one and two dimensions. On the other hand LFs
with $1<\alpha<2$ are recurrent in one dimension but always transient in two
dimensions. Hence in two dimensions LFs will even more significantly reduce
the oversampling of a Brownian searcher. At the same time, however, the
search reliability will go to zero.
In both one and two dimensions (linearly or radially)
LFs are distinct due to the possibility of leapovers, owing to which the
target localization may become less efficient than for Brownian search.
Many search processes indeed fall in the category of (effectively)
one or two dimensions. For example, they are one-dimensional in streams,
along coastlines, or at forest-meadow and other borders. For (relatively)
unbounded search processes as performed by birds or fish, the motion in the
vertical dimension shows a much smaller span than the radial horizontal
motion, and thus becomes effectively two dimensional. If we modify the
condition of blind search and allow the walker to look out for prey while
relocating, in one dimension this would obviously completely change the
picture in favor of LFs with their long unidirectional steps. However,
in two dimensions the radial leapovers would still impede the detection
of the target unless it is exactly crossed during a step.

So what remains of the LF hypothesis? Conceptually, it is
certainly a beautiful idea: a scale-free process reduces oversampling and
thus scans a larger domain. If the target has an extended width, for instance,
a large school instead of a single fish, LFs will then optimize the search
under certain criteria. However, even when the stable index is larger than unity,
in two or three dimensions an LF may also never reach the target, due to its
patchy albeit scale-free exploration of the search space. Thus, even under
LF-friendly conditions such as extremely sparse targets and/or uphill search,
the superiority of LFs over other search models depends on the exact scenario.
For instance, whether it is important that the target is eventually located with
certainty, or whether in an ensemble of equivalent systems only sufficiently many
members need a quick target localization, for instance, the triggering of some
gene expression process responding to a lethal external signal in the cells of a
biofilm.
The LF hypothesis even in the case of blind search without any prior knowledge
is therefore not universal, and depending on the conditions of initial searcher
target separation or the direction of a naturally existing gradient with respect
to the location of the target the regular Brownian motion may be the best search
strategy.

LFs are most efficient under the worst case conditions of blind
search for extremely sparse targets and, as shown here, for uphill motion. While
very rare targets certainly exist in many scenarios, we should qualify the result
for the uphill motion. The above uphill LF scenario holds for abstract processes
such as the blind search of computer algorithms in complex landscapes or for the
topology-mediated LFs in models of gene regulation. For the search of animals
moving against a physical air or water stream, however, we have to take into
consideration that any motion against a gradient requires a higher energy
expenditure. Unless the gradient is very gentle, this aspect relativizes the LF
hypothesis further.

Having said all this, one distinct advantage of spatially scale
free search processes remains. Namely, they are more tolerant to gradually shifting
environmental conditions, for instance, a change in the target distribution, or
when the searcher is exposed to a new patch with conditions unknown to him. This
point is often neglected in the analysis of search processes. A more careful
study of this point may in fact turn back the wheel in favor of LF search.

It should be noted that LFs are processes with a diverging variance  $\left<x^2
(t)\right>$, and may therefore be considered unphysical. There exists the
closely related superdiffusive model of L{\'e}vy walks, they have a finite
variance due to a spatiotemporal coupling introducing a finite travel velocity,
compare, for instance, Ref.~\cite{blumenklafter1987}. This coupling penalizes
long jumps. However, both models converge in the sense that the probability
density function of a L{\'e}vy walker displays a growing L{\'e}vy stable
portion in its center, limited by propagating fronts. The trajectory of such a
L{\'e}vy walk appears increasingly similar to an LF: local search interspersed
by decorrelating long excursions. We expect that at least qualitatively our
present findings remain valid for the case of L{\'e}vy walks. It will be
interesting to investigate this quantitative statement in more detail.

\begin{acknowledgments}
VVP wishes to acknowledge financial support from Deutsche
Forschungsgemeinschaft (Project no.  PA 2042/1-1) as well as discussions
with J. Schulz about simulation of random variables, A. Cherstvy for
help with numerical methods in Mathematica and R. Klages for pointing
out Ref. \cite{Pitchford}. RM acknowledges support from the Academy of
Finland within the FiDiPro scheme. AVCh acknowledges DAAD for financial
support.\end{acknowledgments}

\appendix

\section{Derivation of dimensionless Eq.~(\ref{FFPEdimles})}
\label{dimensionless}
  
We here show how to consistently introduce dimensionless units in the fractional
Fokker-Planck equation. If we denote the dimensionless time and position coordinate
respectively by $\overline{t}$ and $\overline{x}$, such that $t=\overline{t}t_s$
and $x=\overline{x}x_s$ with the dimensional parameters $t_s$ and $x_s$ defined
below. Then we can rewrite Eq.~(\ref{SinkFFPE}) in the form
\begin{equation}
\frac{1}{t_s}\frac{\partial f(x,t)}{\partial\overline{t}}=\frac{K_{\alpha}}{x_s^{
\alpha}}\frac{\partial^{\alpha}f(x,t)}{\partial|\overline{x}|^{\alpha}}-\frac{v}{
x_s}\frac{\partial f(x,t)}{\partial\overline{x}}-\wp_{\mathrm{fa}}(\overline{t}t_s)
\delta(\overline{x}x_s).
\label{A1}
\end{equation}
Note that we are dealing with dimensional density functions so that
\begin{equation}
f(x,t)d(\overline{x}x_s)=\overline{f}(\overline{x},\overline{t})d\overline{x},\\
\end{equation}
and thus
\begin{equation}
f(x,t)=\frac{\overline{f}(\overline{x},\overline{t})}{x_s}.
\end{equation}
Equation (\ref{A1}) then assumes the form
\begin{eqnarray}
\nonumber
\frac{\partial\overline{f}(\overline{x},\overline{t})}{\partial\overline{t}}&=&
\frac{K_{\alpha}t_s}{x_s^{\alpha}}\frac{\partial^{\alpha}\overline{f}(\overline{x},
\overline{t})}{\partial|\overline{x}|^{\alpha}}-\overline{v}\frac{\partial
\overline{f}(\overline{x},\overline{t})}{\partial\overline{x}}-\\
&&-\overline{\wp}_{\mathrm{fa}}(\overline{t})\delta(\overline{x}),
\end{eqnarray}
where $\overline{v}=vt_s/x_s$. We choose the parameters $t_s$ and $x_s$ as
\begin{equation}
x_s=\sigma,\,\,\,t_s=\tau,
\end{equation}
where $\sigma$ and $\tau$ correspond to the scaling factors of the jump length
and waiting time distributions of the continuous time random walk. These appear in the Fourier and Laplace transforms of the jump length and waiting time densities \cite{report}. The respective expansions used in the derivation of the continuous time random walk model for L{\'e}vy flights are
\begin{equation}
\lambda(k)\approx1-\sigma^{\alpha}k^{\alpha},\,\,\,\psi(s)\approx1-s\tau.
\end{equation}
The diffusion coefficient $K_\alpha$ in the fractional Fokker-Planck equation is then expressed in terms of these parameters as \cite{report,enc}
\begin{equation}
K_{\alpha}=\frac{\sigma^\alpha}{\tau}.
\end{equation}
We thus obtain the dimensionless dynamic equation (\ref{FFPEdimles}).

\section{Proof that $P=0$ for $v=0$ and $\alpha\leq1$}
\label{ProofFlat}

To obtain the search reliability $P$ via the relation $P=\wp_{\mathrm{fa}}(s=0)$
one first integrates Eq.~(\ref{pfa}) over $k$ and then takes the limit $s=0$. If $s\neq0$ and $1<\alpha\leq2$, both integrals in the numerator and denominator of Eq.~(\ref{pfa}) converge. In this and the next appendices we prove that $\wp_{\mathrm{fa}}(s)=0$ for $s>0$ and $\alpha\leq1$. Hence $P=0$, which means that the searcher never reaches the target in the case $\alpha\leq1$. 

Taking $v=0$ in Eq.~(\ref{pfa}) we have
\begin{equation}
\wp_{\mathrm{fa}}(s)=\frac{\int_0^{\infty}\cos(kx_0)(s+|k|^{\alpha})^{-1}dk}{\int_0
^{\infty}(s+|k|^{\alpha})^{-1}dk}.
\label{numer}
\end{equation}
For $\alpha\leq1$ the integral in the denominator diverges at infinity. Let us consider the integral in the numerator. We notice that
\begin{equation}
\int_0^{\infty}\frac{\cos(k)}{k^{\alpha}}dk=\Gamma(1-\alpha)\cos(\pi[1-\alpha]),
\label{CosInt}
\end{equation}
for $\alpha<1$. Thus the numerator of expression (\ref{numer}) converges for $\alpha<1$. Since the denominator diverges in this $\alpha$ range, $\wp_{\mathrm{
fa}}(s)=0$ for all values of $s$. Thus, the search reliability vanishes, $P=0$, and the searcher never reaches its target. The limiting case $\alpha=1$ needs separate attention. We observe that the integral (\ref{CosInt}) diverges. For $\alpha=1$,
\begin{equation}
\wp_{\mathrm{fa}}(s)=\lim_{a\to\infty}\frac{\int_0^a\cos(kx_0)(s+k)^{-1}dk}{\int_0
^a(s+k)^{-1}dk}.
\end{equation}
expression for $p_{fa}(s)$. The integrals are
\begin{equation}
\int_0^a(s+k)^{-1}dk=\ln\left(\frac{a+s}{s}\right),
\label{LogInt}
\end{equation}
which logarithmically diverges for $a\rightarrow\infty$, and
\begin{eqnarray}
\nonumber
\int_0^a\frac{\cos(kx_0)}{s+k}dk&=&\int_{sx_0}^{ax_0+sx_0}\frac{\cos(k-s)}{k}dk\\
\nonumber
&&\hspace*{-1.6cm}=\int_{sx_0}^{ax_0+sx}\frac{\cos(k-s)}{k}dk\\
\nonumber
&&\hspace*{-1.6cm}=\cos(s)\int_{sx_0}^{ax_0+sx}\frac{\cos(k)}{k}dk\\
&&\hspace*{-1.2cm}+\sin(s)\int_{sx_0}^{ax_0+sx}\frac{\sin k}{k}dk.
\label{cossin}
\end{eqnarray}
The second term in Eq. (\ref{cossin}) converges. Thus, in the ratio over the divergent integral (\ref{LogInt}) it can be neglected. The first term can be modified to
\begin{eqnarray}
\nonumber
\int_{sx_0}^{ax_0+sx}\frac{\cos(k)}{k}dk&=&\int_{sx_0}^{\infty}\frac{\cos(k)}{k}dk\\
&&-\int_{ax_0+sx}^{\infty}\frac{\cos(k)}{k}dk.
\label{cosint}
\end{eqnarray}
When $a\rightarrow\infty$ the second term in Eq.~(\ref{cosint}) is the cosine
integral at $ax_0+sx$, and it vanishes. Altogether, for any finite $s$
\begin{eqnarray}
\wp_{\mathrm{fa}}(s)=\lim_{a\rightarrow\infty}\frac{\cos(s)\int_{sx_0}^{\infty}
\cos(k)/kdk}{\ln([a+s]/s)}=0,
\end{eqnarray}
which completes the proof.

\section{Proof that $P=0$ for $v\neq0$ and $\alpha\leq1$}
\label{ProofBias}

Let us start from the Cauchy case $\alpha=1$. The expression for the first arrival density follows from Eq.~(\ref{pfa}),
\begin{equation}
\wp_{\mathrm{fa}}(s)=\frac{\int_{-\infty}^{\infty}\{\cos(kx_0[s+|k|])-vk\sin(kx_0)
\}\overline{\beth}dk}{\int_{-\infty}^{\infty}(s+|k|)\overline{\beth}dk}
\label{alpha1ratio}
\end{equation}
with
\begin{equation}
\overline{\beth}=\frac{1}{(s+|k|)^2+k^2v^2}.
\end{equation}
Alternatively,
\begin{equation}
\wp_{\mathrm{fa}}(s)=\lim_{a\rightarrow\infty}\frac{\int_0^a\{\cos(kx_0[s+|k|])
-vk\sin(kx_0)\}\overline{\beth}dk}{\int_0^a(s+|k|)\overline{\beth}dk}.
\label{C2}
\end{equation}
Let us first consider the integral in the denominator,
\begin{widetext}
\begin{equation}
\int_0^a(s+|k|)\overline{\beth}dk=\frac{1}{2(1+v^2)}\ln\frac{(1+v^2)a^2+2sa+s^2}{
s^2}-\left(\arctan\frac{s+(1+v^2)a}{|v|s}-\arctan\frac{1}{|v|}\right)\left(\frac{1
}{|v|}+\frac{1}{(1+v^2)|v|}\right).
\end{equation}
\end{widetext}
At $a\rightarrow\infty$ only the first term is significant, and it diverges
logarithmically. The integral in the numerator converges due to the oscillating
functions in the integrands. With the diverging denominator and the converging
numerator in Eq.~(\ref{C2}), we have that $\wp_{\mathrm{fa}}(s)=0$ for any finite $s$. The search reliability vanishes.

For $\alpha<1$ in Eq.~(\ref{pfa}) with $k\rightarrow\infty$ we have $k^\alpha\ll
k$, and the proof is analogous to the case just considered.

\section{Solution for $v=0$ via Fox $H$-functions}
\label{HfuncV0}

Without a bias, Eq.~(\ref{pfa}) takes on the form
\begin{equation}
\wp_{\mathrm{fa}}(s)=\frac{\int_0^{\infty}\cos(k)\aleph dk}{\int_0^{
\infty}\aleph dk}\equiv\frac{I_2}{I_1}.
\end{equation}
where the abbreviation $\aleph$ is defined in Eq.~(\ref{aleph}).
The integral in the denominator yields \cite{Prudnikov}
\begin{equation}
I_1=\frac{1}{\alpha}\left(\frac{1}{sx_0^{\alpha}}\right)^{(\alpha-1)/\alpha}\Gamma
\left(\frac{1}{\alpha}\right)\Gamma\left(1-\frac{1}{\alpha}\right).
\label{v0denom}
\end{equation}
The integral in the numerator can be obtained in terms of the Fox $H$-function
technique \cite{Saxena2010}. Since
\begin{equation}
\frac{1}{1+x^{\alpha}}=H^{11}_{11}\left[x\left|\begin{array}{l}(0,1/\alpha)\\
(0,1/\alpha)\end{array}\right.\right],
\label{xalphaplusone}
\end{equation}
we find that
\begin{widetext}
\begin{equation}
I_2=\left(\frac{1}{sx_0^{\alpha}}\right)^{(\alpha-1)/\alpha}\int^{\infty}_0\frac{
\cos\left[s^{1/\alpha}x_0y\right]}{1+y^{\alpha}}dk=\frac{\sqrt{\pi}}{\alpha}\frac{
1}{sx_0^\alpha}H^{12}_{31}\left[\frac{2}{s^{1/\alpha}x_0}\left|\begin{array}{l}
(1/2,1/2),(0,1/\alpha),(0,1/2)\\(0,1/\alpha)\end{array}\right.\right],
\label{cosnumer}
\end{equation}
where we used Eq.~(\ref{xalphaplusone}) and the integral (2.25.2.4) from
Ref.~\cite{Prudnikov}. Transforming the $H$-function by help of the properties
(1.3) and (1.5) from Ref.~\cite{Saxena2010} we obtain
\begin{equation}
\wp_{\mathrm{fa}}(s)=\frac{\sqrt{\pi}}{2\Gamma(1/\alpha)\Gamma(1-1/\alpha)}H^{21}_{
13}\left[\frac{1}{2}s^{1/\alpha}x_0\left|\begin{array}{l}([\alpha-1]/\alpha,1/\alpha
)\\(0,1/2),([\alpha-1]/\alpha,1/\alpha),(1/2,1/2)\end{array}\right.\right],
\label{pfaSolutionFlat}
\end{equation}
Using the properties of the Laplace transform of the $H$-function (see chapter 2 in Ref.~\cite{Saxena2010}) we get the first arrival density in the time domain,
\begin{eqnarray}
\wp_{\mathrm{fa}}(t)=\frac{\alpha\sqrt{\pi}}{2\Gamma(1/\alpha)\Gamma(1-1/\alpha)t}
H^{21}_{23}\left[\frac{x_0^{\alpha}}{2^{\alpha}t}\left|\begin{array}{l}(1/2,1),
(0,1)\\(0,\alpha/2),([\alpha-1]/\alpha,1),(1/2,\alpha/2)\end{array}\right.\right].
\label{pfatimedomain}
\end{eqnarray}
Expansion of Eq.~(\ref{pfatimedomain}) in the long-time limit yields
Eq.~(\ref{asymp}), where
\begin{equation}
C(\alpha)=\frac{\alpha\sin^2(\pi/\alpha)\sin(\pi[2-\alpha]/2)\Gamma(2-\alpha)
\Gamma(2-1/\alpha)}{\pi^2(\alpha-1)}
\label{Calpha}.
\end{equation}
\end{widetext}

\section{General solution for the Brownian case}
\label{BrownDeriv}

It is instructive to obtain the well-known first arrival density in the Brownian
case directly from Eq.~(\ref{pfa}). For $\alpha=2$, Eq.~(\ref{pfa}) assumes the form
\begin{equation}
\wp_{\mathrm{fa}}(s)=\frac{\int_{-\infty}^{\infty}\exp(ikx_0)\overline{\daleth}dk}{
\int_{-\infty}^{\infty}\overline{\daleth}dk}\equiv\frac{I_1}{I_2},
\label{brown}
\end{equation}
where
\begin{equation}
\overline{\daleth}=\frac{1}{s+K_2k^2-ikv}
\end{equation}
The denominators in these integrals are quadratic polynomials in $k$ and hence can be rewritten as $K_2(k-k_1)(k-k_2)$, where
\begin{equation}
k_{1,2}=\frac{iv}{2K_2}\pm i\sqrt{\frac{v^2}{4K_2^2}+\frac{s}{K_2}}.
\end{equation}
Then both integrals can be easily calculated by the method of residues, and we arrive at Eq.~(\ref{pfabrownbias}).

At short times ($v^2/K_2<s$) Eq.~(\ref{pfabrownbias}) yields
\begin{equation}
\wp_{\mathrm{fa}}(s)=(1-\mathrm{Pe}_{2})\exp\left(-\sqrt{\frac{sx_0^2}{K_2}}\right).
\label{pfaApproxBrownian}
\end{equation} 
This result can be obtained by first expanding Eq.~(\ref{brown}) at small
P{\'e}clet numbers and then integrating each of the terms over $k$. Indeed,
from Eq.~(\ref{brown}) we get
\begin{equation}
\wp_{\mathrm{fa}}(s)\simeq\frac{\int_{-\infty}^{\infty}\cos(k)\tilde{\beth}dk-
\int_{-\infty}^{\infty}2\mathrm{Pe}_2k\sin(k)\tilde{\beth}^2dk}{\int_{-\infty}^{
\infty}\tilde{\beth}dk}
\label{pfaPecletBr}
\end{equation}
with
\begin{equation}
\tilde{\beth}=\frac{1}{sx_0^2/K_2+k^2}
\end{equation}
The three integrals in Eq.~(\ref{pfaPecletBr}) become
\begin{eqnarray}
\nonumber
&&\int_{-\infty}^{\infty}\tilde{\beth}dk=\pi\sqrt{\frac{K_2}{sx_0^2}},\\
\nonumber
&&\int_{-\infty}^{\infty}\cos(k)\tilde{\beth}dk=\pi\sqrt{\frac{K_2}{sx_0^2}}\exp
\left(-\sqrt{\frac{sx_0^2}{K_2}}\right),\\
\nonumber
&&\int_{-\infty}^{\infty}2\mathrm{Pe}_2k\sin(k)\tilde{\beth}^2dk=\pi\mathrm{Pe}_2
\sqrt{\frac{K_2}{sx_0^2}}\exp\left(-\sqrt{\frac{sx_0^2}{K_2}}\right),
\end{eqnarray}
which yields the result (\ref{pfaApproxBrownian}).

Finally, we note that the inverse Laplace transform of Eq.~(\ref{pfabrownbias})
leads to the expression in time domain,
\begin{equation}
\wp_{\mathrm{fa}}(t)=\frac{x_0}{\sqrt{4\pi t^3}}\exp\left(-\frac{(vt+x_0)^2}{4t}
\right).
\label{alpha2}
\end{equation}
This result coincides with the solution obtained by either the Green's function
technique or the images method in Ref.~\cite{Redner} (see Eq. (3.2.13) there).

\section{Expansion (\ref{pfaexp}) in terms of $H$-functions}
\label{AppGenExp}

We show here how expansion (\ref{pfaexp}) is obtained in terms of $H$-functions.
Two out of three integrals in Eq.~(\ref{pfaexp}) were computed above in
App.~\ref{HfuncV0} as Eqs.~(\ref{v0denom}) and (\ref{cosnumer}). The last unknown integral from expression (\ref{pfaexp}) can be computed in a similar way,
\begin{widetext}
\begin{eqnarray}
\nonumber
\int_{-\infty}^{\infty}\frac{2\mathrm{Pe}_{\alpha}k\sin k}{\left(sx_0^{\alpha}+|k|
^{\alpha}\right)^2}dk&=&4\mathrm{Pe}_{\alpha}\left(\frac{1}{sx_0^\alpha}\right)^{2
-2/\alpha}\int_0^{\infty}\frac{y\sin\left(s^{1/\alpha}x_0y\right)}{\left(1+y^{
\alpha}\right)^2}dy\\
\nonumber
&=&\frac{4\mathrm{Pe}_{\alpha}}{\alpha}\left(\frac{1}{sx_0^\alpha}\right)^{2-2/
\alpha}\int_0^{\infty}\sin\left(s^{1/\alpha}x_0y\right)yH^{11}_{11}\left[y\left|
\begin{array}{l}(-1,1/\alpha)\\(0,1/\alpha)\end{array}\right.\right]dy\\
&=&\frac{8\mathrm{Pe}_{\alpha}\sqrt{\pi}}{\alpha}\left(sx_0^{\alpha}\right)^2
H^{12}_{31}\left[2s^{1/\alpha}x_0\left|\begin{array}{l}(-1/2,1/2),(-1,1/\alpha),
(0,1/2)\\(0,1/\alpha)\end{array}\right.\right].
\end{eqnarray}
With these results we obtain the following expression in Laplace space,
\begin{eqnarray}
\nonumber
\wp_{\mathrm{fa}}(s)&=&\frac{\sqrt{\pi}}{2\Gamma(1/\alpha)\Gamma(1-1/\alpha)}
\left(H^{12}_{31}\left[\frac{2}{s^{
1/\alpha}x_0}\left|\begin{array}{l}(1,1/2),(1/\alpha,1/\alpha),(1/2,1/2)\\
(1/\alpha,1/\alpha)\end{array}\right.\right]\right.\\
&&\left.-2^{{2-\alpha}}\mathrm{Pe}_{\alpha}H^{12}_{31}\left[\frac{2}{s^{1/\alpha}
x_0}\left|\begin{array}{l}(\alpha/2,1/2),(1/\alpha,1/\alpha),([\alpha+1]/2,1/2)\\
([\alpha+1]/\alpha,1/\alpha)\end{array}\right.\right]\right).
\label{pfaExpansionGen}
\end{eqnarray}
Inverse Laplace transform of Eq.~(\ref{pfaExpansionGen}) yields
\begin{eqnarray}
\nonumber
\wp_{\mathrm{fa}}(t)&=&\frac{\alpha\sqrt{\pi}}{2\Gamma(1/\alpha)\Gamma(1-1/\alpha)t}
\left(H^{21}_{23}\left[
\frac{x_0^{\alpha}}{2^{\alpha}t}\left|\begin{array}{l}(1/2,1),(0,1)\\(0,\alpha/2),
([\alpha-1]/\alpha,1),(1/2,\alpha/2)\end{array}\right.\right]\right.\\
&&\left.-\mathrm{Pe}_{\alpha}H^{21}_{23}\left[\frac{x_0^{\alpha}}{2^{\alpha}t}\left|
\begin{array}{l}(-1/\alpha,1),(0,1)\\(1-\alpha/2,\alpha/2),([\alpha-1]/\alpha,1),
([1-\alpha]/2,\alpha/2)\end{array}\right.\right]\right).
\label{pfaExpansionGenT}
\end{eqnarray}
\end{widetext}

\section{Derivation of the Brownian weak bias expansion (\ref{pfaExpansionGen})}
\label{BrownLimit}

We represent the first arrival density $\wp_{\mathrm{fa}}$ from
Eq.~(\ref{pfaExpansionGen}) as $\wp_{\mathrm{fa}}=\wp_{\mathrm{fa}}^{(1)}+
\wp_{\mathrm{fa}}^{(2)}$, where the first and second contribution correspond to the first and second terms in the expression (\ref{pfaExpansionGen}). Then for $\alpha=2$ the order of $H$-function is reduced by use of the properties 1.2 and 1.3 from chapter 1 in Ref.~\cite{Saxena2010}) as well as its definition via the Mellin transform \cite{Saxena2010}. This procedure yields
\begin{widetext}
\begin{eqnarray}
\nonumber
\wp_{\mathrm{fa}}^{(1)}(s)&=&\frac{1}{2\sqrt{\pi}}H^{12}_{31}\left[\frac{2K_2^{1/2}}
{s^{1/2}x_0}\left|\begin{array}{l}(1,1/2),(1/2,1/2),(1/2,1/2)\\[0.2cm](1/2,1/2)
\end{array}\right.\right]=\frac{1}{2\sqrt{\pi}}H^{02}_{20}\left[\frac{2K_2^{1/2}}
{s^{1/2}x_0}\left|\begin{array}{l}(1,1/2),(1/2,1/2)\\[0.2cm]\rule{1.2cm}{0.02cm}
\end{array}\right.\right]=\\
&=&\frac{1}{2\sqrt\pi}H^{20}_{02}\left[\frac{s^{1/2}x_0}{2K_2^{1/2}}\left|
\begin{array}{l}\rule{1.2cm}{0.02cm}\\[0.2cm](0,1/2),(1/2,1/2)\end{array}\right.
\right]=H^{10}_{01}\left[\frac{s^{1/2}x_0}{K_2^{1/2}}\left|\begin{array}{l}
\rule{1.2cm}{0.02cm}\\[0.2cm](0,1)\end{array}\right.\right]=\exp\left(-
\frac{s^{1/2}{x_0}}{K_2^{1/2}}\right).
\end{eqnarray}
Similar steps for $\wp_{\mathrm{fa}}^{(2)}$ lead to the result
\begin{eqnarray}
\wp_{\mathrm{fa}}^{(2)}=-\frac{\mathrm{Pe}_2}{2\sqrt{\pi}}H^{12}_{31}\left[\frac{
2}{z}\left|\begin{array}{l}(1,1/2),(1/2,1/2),(3/2,1/2)\\[0.2cm](3/2,1/2)\end{array}
\right.\right]=-\mathrm{Pe}_2\exp\left(-\frac{s^{1/2}{x_0}}{K_2^{1/2}}\right).
\end{eqnarray}
Thus $\wp_{\mathrm{fa}}=(1-\mathrm{Pe}_2)\exp\left(-s^{1/2}x_0/K_2^{1/2}\right)$,
which is the expansion of the general solution in the Brownian case, expression
(\ref{pfaApproxBrownian}). The same result can be obtained by calculations
in $t$-space
\begin{eqnarray}
\wp_{\mathrm{fa}}^{(1)}=\frac{1}{\sqrt{\pi}t}H^{21}_{23}\left[\frac{x_0^{2}}{4K_2t}
\left|\begin{array}{l}(1/2,1),(0,1)\\[0.2cm](0,1),(1/2,1),(1/2,1)\end{array}\right.
\right]=\frac{x_0}{\sqrt{4\pi t^3}}\exp\left(-\frac{x_0^2}{4K_2t}\right),
\end{eqnarray}
and
\begin{eqnarray}
\wp_{\mathrm{fa}}^{(2)}=-\frac{\mathrm{Pe}_2}{\sqrt{\pi}t}H^{21}_{23}\left[\frac{
x_0^{2}}{4K_2t}\left|\begin{array}{l}(-1/2,1),(0,1)\\[0.2cm](0,1),(1/2,1),(-1/2,1)
\end{array}\right.\right]=-\frac{\mathrm{Pe}_2 x_0}{\sqrt{4\pi t^3}}\exp\left(-
\frac{x_0^2}{4K_2t}\right).
\end{eqnarray}

\section{Implicit Fox $H$-function expression for $\wp_{\mathrm{fa}}(s)$}
\label{implicitApp}

The expressions for $H_1(t)$ and $H_2(t)$ in Eq.~(\ref{alter}) can be obtained
by help of standard properties of $H$-function \cite{Prudnikov} and the
identification for the exponential function,
\begin{equation}
e^{-z}=H^{10}_{01}\left[z\left|\begin{array}{l}\rule{0.8cm}{0.02cm}\\[0.2cm](0,1)
\end{array}\right.\right].
\end{equation}
Consequently,
\begin{eqnarray}
H_1(t)=\int_0^\infty\cos(k|vt|)H^{10}_{01}\left[tk^{\alpha}\left|\begin{array}{l}
\rule{0.8cm}{0.02cm}\\[0.2cm](0,1)\end{array}\right.\right]dk
=\frac{\sqrt{\pi}}{|vt|}H^{11}_{21}\left[t\left(\frac{2}{|vt|}\right)^{\alpha}
\left|\begin{array}{l}(1/2,\alpha/2),(0,\alpha/2)\\[0.2cm](0,1)\end{array}\right.
\right]
\end{eqnarray}
and
\begin{eqnarray}
H_2(t)=\int_0^\infty\cos(k|vt+x_0|)\exp\left(-kt^{\alpha}\right)dk=
\frac{\sqrt{\pi}}{|vt+x_{0}|}H^{11}_{21}\left[\left(\frac{2}{
|vt+x_{0}|}\right)^{\alpha}t\left|\begin{array}{l}(1/2,\alpha/2),(0,\alpha/2)\\
[0.2cm](0,1)\end{array}\right.\right].
\label{H2tApp}
\end{eqnarray}
To construct the expression (\ref{ratio}) for the first arrival density, we need the Laplace transforms of the functions $H_i(t)$. For $H_1(s)$ we find
\begin{eqnarray}
\nonumber
H_1(s)&=&\mathcal{L}\left\{\frac{\sqrt{\pi}}{|v|\tau}H^{11}_{12}\left[\tau^{\alpha
-1}\left(\frac{|v|}{2}\right)^{\alpha}\left|\begin{array}{l}(1,1)\\[0.2cm](1/2,
\alpha/2),(1,\alpha/2)\end{array}\right.\right]\right\}\\
&=&\frac{\sqrt{\pi}}{2}s^{1/\alpha-1}H^{12}_{22}\left[s^{1-\alpha}\left(\frac{|v|}{
2}\right)^{\alpha}\left|\begin{array}{l}(1/\alpha,\alpha-1),(1-1/\alpha,1)\\[0.2cm]
(0,\alpha/2),(1/2,\alpha/2)\end{array}\right.\right]
\end{eqnarray}
Using the expansion of the $H$-function at small arguments \cite{Saxena2010} we find at $v=0$
\begin{equation}
H_1(s)=\frac{s^{1/\alpha-1}}{\alpha}\Gamma\left(1-\frac{1}{\alpha}\right)\Gamma
\left(\frac{1}{\alpha}\right),
\end{equation}
which is exactly the same result as one can get by direct computation of the integral $\int_0^{\infty}e^{-k^\alpha t}dk$ and subsequent Laplace transform.

At $v=0$ from Eq.~(\ref{H2tApp}) we get by direct Laplace transform
\begin{eqnarray}
H_2(s)=\frac{\sqrt{\pi}}{x_0s}H^{12}_{31}\left[\frac{2^{\alpha}}{sx_0^{\alpha}}
\left|\begin{array}{l}(0,1),(1/2,\alpha/2),(0,\alpha/2)\\[0.2cm](0,1)\end{array}
\right.\right]
=\frac{\sqrt{\pi}}{\alpha x_0s}H^{12}_{31}\left[\frac{2}{s^{1/\alpha}x_0}\left|
\begin{array}{l}(0,1/\alpha),(1/2,1/2),(0,1/2)\\[0.2cm](0,1/\alpha)\end{array}
\right.\right]
\end{eqnarray} 
and hence
\begin{eqnarray}
\nonumber
\wp_{\mathrm{fa}}(s)&=&\frac{H_2(s)}{H_1(s)}=\frac{\sqrt{\pi}}{x_0s^{1/\alpha}
\Gamma(1/\alpha)\Gamma(1-1/\alpha)}
H^{12}_{31}\left[\frac{2}{s^{1/\alpha}x_0}\left|\begin{array}{l}(0,1/
\alpha),(1/2,1/2),(0,1/2)\\[0.2cm](0,1/\alpha)\end{array}\right.\right]\\
&=&\frac{\sqrt{\pi}}{2\Gamma(1/\alpha)\Gamma(1-1/\alpha)}
H^{12}_{31}\left[\frac{2}{s^{1/\alpha}x_0}\left|
\begin{array}{l}(1/\alpha,1/\alpha),(1,1/2),(1/2,1/2)\\[0.2cm](1/\alpha,1/\alpha)
\end{array}\right.\right].
\end{eqnarray}
We see that this expression is different from Eq.~(\ref{pfaSolutionFlat}) in the order of the first two brackets in the top row of the $H$-function. However, these brackets can be exchanged due to property 1.1 of the $H$-function in Ref.~\cite{Saxena2010}. Thus, the $H$-function solution for the unbiased case ($v=0$) is obtained correctly.

Now let us derive the result for any $v$ in the limit $\alpha=2$. For that
purpose we note that
\begin{eqnarray}
H_2(t)=\frac{\sqrt{\pi}}{2}t^{1/\alpha}H^{11}_{12}\left[\frac{1}{t}\left(\frac{
|vt+x_{0}|}{2}\right)^{\alpha}\left|\begin{array}{l}(1-1/\alpha,1)\\[0.2cm]
(0,\alpha/2),(1/2,\alpha/2)\end{array}\right.\right].
\end{eqnarray}   
For $\alpha=2$ the reduction formula for $H$-functions (property 1.2
of Ref.~\cite{Saxena2010}) yields
\begin{equation}
H_2(t)=\frac{\sqrt{\pi}}{2\sqrt{t}}\exp\left(\frac{(vt+x_0)^2}{4K_2t}\right),
\end{equation}
where we restored the Brownian diffusivity. Alternatively,
\begin{equation}
H_2(s)=\frac{\pi}{\sqrt{4s+v^2}}\exp\left(-\frac{1}{2}x_0\left[v+\sqrt{4s+v^2}
\right]\right)
\end{equation}
Since $H_1(s)=H_2(s)|_{x_0=0}$,
\begin{equation}
\wp_{\mathrm{fa}}(s)=\exp\left(-\frac{1}{2}x_0\left[v+\sqrt{4s+v^2}\right]\right).
\end{equation}
Inverse Laplace transform of the latter relation produces Eq.~(\ref{alpha2}).
\end{widetext}

\end{document}